%% file: Main.tex
\documentclass[lettersize,journal,twocolumn]{IEEEtran}

\usepackage{amsmath,amsfonts}
\usepackage{algorithmic}
\usepackage{algorithm}
\usepackage{array}
\usepackage[caption=false,font=normalsize,labelfont=sf,textfont=sf]{subfig}
\usepackage{textcomp}
\usepackage{stfloats}
\usepackage{url}
\usepackage{verbatim}
\usepackage{graphicx}
\usepackage{cite}
\input{macros}

\usepackage[font=small,labelfont=bf,tableposition=top]{caption}
\title{G-SemTMO: Tone Mapping with a Trainable Semantic Graph}
\author{Abhishek Goswami, Erwan Bernard, Wolf Hauser, Frederic Dufaux, \IEEEmembership{Fellow, IEEE} and Rafal Mantiuk, \IEEEmembership{Member, IEEE}
\thanks{A. Goswami is with DxO Labs and L2S, Université Paris-Saclay, CNRS, CentraleSupélec, France. (email: abhishek.goswami@centralesupelec.fr)}
\thanks{E. Bernard and W. Hauser are with DxO Labs, France. (email: \{ebernard, whauser\}@dxo.com)}
\thanks{F. Dufaux is with the L2S, Université Paris-Saclay, CNRS, CentraleSupélec, France. (email: frederic.dufaux@centralesupelec.fr)}
\thanks{R. Mantiuk is with the Department of Computer Science and Technology, University of Cambridge, UK. (email: rkm38@cam.ac.uk)}
}

\begin{document}
\maketitle

\begin{abstract}
A Tone Mapping Operator (TMO) is required to render images with a High Dynamic Range (HDR) on media with limited dynamic capabilities. TMOs compress the dynamic range with the aim of preserving the visually perceptual cues of the scene. Previous literature has established the benefits of TMOs being semantic aware, understanding the content in the scene to preserve the cues better. Expert photographers analyze the semantic and the contextual information of a scene and decide tonal transformations or local luminance adjustments. This process can be considered a manual analogy to tone mapping. In this work, we draw inspiration from an expert photographer's approach and present a Graph-based Semantic-aware Tone Mapping Operator, G-SemTMO. We leverage semantic information as well as the contextual information of the scene in the form of a graph capturing the spatial arrangements of its semantic segments. Using Graph Convolutional Network (GCN), we predict intermediate parameters called Semantic Hints and use these parameters to apply tonal adjustments locally to different semantic segments in the image. In addition, we also introduce LocHDR, a dataset of 781 HDR images tone mapped manually by an expert photo-retoucher with local tonal enhancements. We conduct ablation studies to show that our approach, G-SemTMO\footnote{Code and dataset to be published with the final version of the manuscript}, can learn both global and local tonal transformations from a pair of input linear and manually retouched images by leveraging the semantic graphs and produce better results than both classical and learning based TMOs. We also conduct ablation experiments to validate the advantage of using GCN.
\end{abstract}

\begin{IEEEkeywords}
HDR Imaging, Tone Mapping Operators, Graph Convolutional Networks, Semantic Awareness, Deep Learning.
\end{IEEEkeywords}

\section{Introduction}
Tone mapping operators compress the dynamic range of an image, trying to preserve its aesthetic and visual quality. 
The problem of finding a balance between dynamic range compression and aesthetic quality predates digital image processing. Renaissance painters created high fidelity paintings with the limited dynamic range of pigments while maintaining contextual cues of the scene. In the era of analog photography, photo-retouchers reproduced high dynamic range content on limited dynamic media by locally adjusting exposure and contrast~\cite{ansel1983print}. Artists naturally took image semantics into account in order to reproduce the visual cues of the scene. Therefore, while a TMO maps the luminance values from a linear image to its output, it helps if it is also aware of the content in the scene. 

The importance of TMOs being aware of the semantic context of a scene has been established in literature \cite{goswami2020tone}. The specific point of interest lies in the question, how can we use contextual semantic information explicitly in the tone mapping pipeline? We hypothesize that ideally a TMO should analyse an image like an expert photographer, generate an abstract understanding of the scene and modify the image locally based on the abstract semantic information. \\
\textit{How do photographers analyse a scene while retouching?} Parsing a scene is essential to aesthetically modify an image. Learning-based semantic segmentation networks assign fine-grained labels to pixels and generate a semantic map for an image~\cite{wu2019fastfcn}. Unlike the fine semantic segmentation, photographers parse a scene on a coarser level. They identify photographically important objects in the scene. The abstract information combining the semantic labels, the context and the global attributes of the labels, such as the luminance distribution, play a significant role in deciding the local enhancement. We call these abstracted semantic information - \textit{`Semantic Hints'}. These hints drive the local adjustment of tonal values. In summary, in this work, we propose:  
\begin{itemize}
    \item a tone mapping operator which learns the tonal transformation as a function of semantic and contextual information of the image.
    \item a GCN to exploit the semantic information from the spatial arrangement of semantic segments in the image and predict aforementioned semantic hints.
    \item to exploit the hints in conjunction with the semantic features from the linear image to predict a tone mapped image aesthetically and perceptually close to a retouched version as generated by an expert photographer.
    \item LocHDR, a locally enhanced dataset of 781 HDR images tone mapped manually by an expert photo-retoucher.

\end{itemize}

\section{Related Work}
\IEEEpubidadjcol
The term ``tone mapping" is used to describe a broad range of techniques, often solving different problems. Therefore, it is important to position our research in that broader scope. The three main application areas of tone mapping are computer graphics, HDR video/television, and photography. In computer graphics tone mapping is used in the final stages of the rendering pipeline to simulate either a camera or the eye. Tone-mapping in graphics is often intended to bring a cinematographic look by simulating lens softness, or flare \cite{Hullin2011}. Alternatively, it could be used to mimic the appearance of scene as it would be perceived by the eye, for example, by simulating night vision \cite{Irawan2005,Wanat2014}. Another important application is HDR video and television, where color graded HDR content needs to be mapped to a display that may offer lower dynamic range and brightness than the reference HDR display used for color grading \cite{SMPTE2017}. This paper focuses on the application of tone mapping in photography, where the goal is to produce images of certain aesthetics from linear (RAW) images captured by a camera sensor. All three application areas are not mutually exclusive, however, their input and aims are distinct. 

The early tone mapping techniques for photography relied on heuristics or rules often inspired by photographic practices, intended to reproduce images of good contrast on displays of limited dynamic range \cite{reinhard2002photographic,Krawczyk05EG}. Later methods were guided by optimization, which attempted to find the best reproduction by minimizing a perceptual difference between the original and reproduced images \cite{mantiuk2008display,KedeMa2015}. More recently, machine learning was introduced to tone mapping as a tool to learn mapping from RAW/HDR/linear images to their desired reproduction from a large dataset of training examples~\cite{fivek}. Since the main goal of photographic tone mapping is reproducing loosely defined image aesthetics, the problem is an excellent fit for  machine learning techniques, which can learn from a large number of examples, without the need for well-defined rules. 

Tone mapping can be considered as a regression problem, in which the goal is to learn a function mapping from input HDR, RAW or linear images to the desired tone mapped images, usually provided by a large dataset of input/output examples. Such regression could be realized by standard techniques, such as LASSO (least absolute shrinkage and selection operator) or GPR (gaussian process regression) \cite{fivek}, by finding nearest-neighbors in a dataset of reference images \cite{hwang2012context}, using a fully connected neural network to learn the coefficients of the quadratic polynomial basis functions \cite{yan2016automatic}, or learning simple brightness adjustment for semantic segments \cite{goswami2020tone}. More recent methods involve a combination of fully connected and convolutional neural networks to extract both local and global (contextual) features from images \cite{gharbi2017deep}. Another popular choice is encoder-decoder architecture, based on convolutional neural networks \cite{Montulet2019,Rana2020}. One common feature in all these methods is that the input to the regression typically combines local features, such as pixel color and its neighborhood and global features, such as image statistics, contextual or semantic information. Our method expands on this concept by explicitly modeling a trainable semantic graph of the image, which guides the tone mapping process.

All the aforementioned learning based methods implicitly use semantic information in different forms to improve tone mapping. However, we realise that semantic awareness is not limited to learning local or global attributes based on semantic categories. It also involves analysing the context under which the semantic categories are observed. Hence, we explicitly analyse semantic information through a graph of connected semantic segments. A Graph Convolutional Network (GCN) helps us pass information between nodes in the graph \cite{kipf2017semisupervised} and learn local adjustments based on contextual information. 
A comparative study of graph neural networks and its applications \cite{zhou2020graph} lists the domain of computer vision and image sciences where GCN has been applied for image classification \cite{wang_classification, kampffmeyer2019rethinking}, segmentation \cite{liang2016semantic} and reasoning \cite{wang2018deep}. However, to the best of our knowledge, our work is the first attempt to apply GCN as a model of trainable image semantics for tone mapping. Although digital images have a regular grid-like structure, their semantic segmentation maps combined with attributes per segment leads to an irregular data structure fit for graph-based representation.

Training a GCN to learn contextual information from semantic categories and how it affects tonal modification, requires a dataset of input-and-retouched image pairs. MIT Adobe FiveK \cite{fivek} dataset provides 5000 RAW images and their retouched versions created manually by 5 expert photographers. This dataset has been used to learn expert retouching styles, most notably for HDRNET \cite{gharbi2017deep}. 

\begin{figure}[]
\centering
  \includegraphics[width=.74\linewidth]{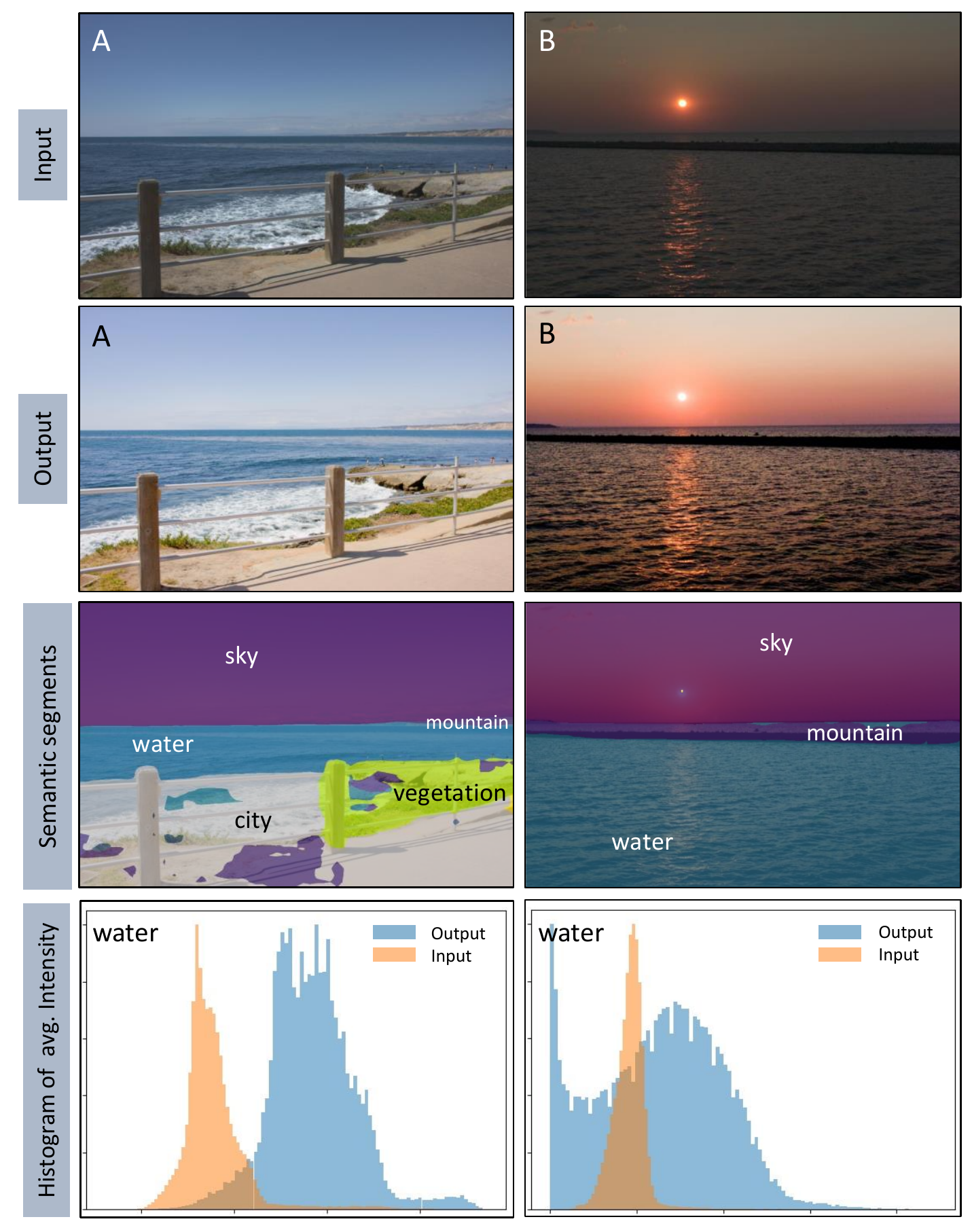}
  \caption{Understanding semantic awareness. \textit{Row 1}: Gamma corrected input images A (\textit{a1824}) and B (\textit{a1892}). \textit{Row 2}: Images manually retouched by expert E from MIT Adobe FiveK \cite{fivek}. \textit{Row 3}: Coarse semantic segments --- fine labels obtained via FastFCN \cite{wu2019fastfcn} segmentation and merged as per SemanticTMO \cite{goswami2020tone}. \textit{Bottom}: Input and output average intensity histograms for the `\textit{water}' semantic segment. Histograms show markedly different output distribution for relatively similar input distribution.} 
  \label{fig:semantic-awareness}
\end{figure}

\section{Semantic-aware tone mapping}
\label{sec:pipeline desc}
We propose a neural network architecture that is trained over the pairs of RAW linear and expert retouched images. The network learns to generate latent hints based on the semantic content of the image and adjust tone mapping based on this semantic information.
In the following subsections we describe our new learning-based pipeline. The pipeline has broadly two modules: a \textit{Semantic Hints Module} and a \textit{Tone Mapping Module}. The semantic hints module drives the semantic awareness of the TMO and generates the aforementioned hints. The application module works as a $n$-dimensional lookup table and learns a mapping as a function of the aforementioned hints. Before we dive deeper, it is necessary to to delve into the notion of semantic awareness.

\subsection{Introducing Semantic Awareness}

To introduce semantic awareness, we incorporate the semantic features of a scene, based on the different labels obtained using semantic segmentation algorithm, \textit{e.g.} the color and luminance statistics per semantic label. We also incorporate the contextual understanding of the scene through a graph representing the neighborhood and spatial arrangements of the semantic labels in the segmentation map. 
We hypothesize that, along with the semantic features, the node-level neighborhood semantic information guides the image enhancement while retouching images.

\figref{semantic-awareness} shows two images \textit{A} and \textit{B} from the Adobe FiveK \cite{fivek} dataset, both manually retouched by expert E. We use FastFCN semantic segmentation algorithm \cite{wu2019fastfcn} and merge the labels to coarser bins as suggested in SemanticTMO \cite{goswami2020tone}. Although visually both images have a similar composition, semantic decomposition reveals the difference in semantic labels and their neighbors. The \textit{water} semantic segment is surrounded by \textit{sky} and \textit{mountain} in image \textit{B}, whereas in image \textit{A} the \textit{vegetation} and \textit{city} are also neighbors to \textit{water}. For both images, we plot the intensity histograms of the \textit{water} segments for both the gamma-corrected input image and output image modified by the expert. The input histograms have a similar narrow distribution, although visibly shifted to the left for image \textit{B} due to the overall low light. However, the output histograms show a very different distribution. The two segments receive different tonal adjustments despite having the same semantic label. This prompts us to conclude that the tonal adjustments are not just a function of semantic-based priors, but also of the local neighborhood of the semantic labels and their attributes, such as the intensity distribution or label information.

Hence, we propose a learning-based tone mapping algorithm which leverages spatial semantic information, as well as the contextual information in the form of a graph capturing the spatial arrangements of the segments. 

\subsection{Semantic Hint Module}
\label{sec:semantic_hint_module}
An image can be segmented into several regions of semantic consistency by a semantic segmentation network. The segmentation map can be represented as a connected graph in which each node corresponds to a semantic segment and an edge is inserted when two semantic segments are neighbors in the map. This representation should mimic the way a photographer may analyze the semantic information in an image. 

Formally, an input image $I$ with linear color values and with $n$ semantic segments can be represented as a graph $\mathcal{G}=(\mathcal{V}, \mathcal{E})$ where $\mathcal{V}$ are $n$ nodes corresponding to the segments, and $\mathcal{E}$ are the edges, represented as an adjacency matrix, such that $\mathcal{E}_{i,j}=1$ if the segments corresponding to the nodes $i$ and $j$ are neighbours to each other. A \textit{Graph Convolutional Network} (GCN) \cite{kipf2017semisupervised} is trained to learn a function on the graph $\mathcal{G}$. More specifically, it takes, for each node in the graph, an input feature vector $\textbf{x}_i, i\in n$, summarised in a $n\times d$ feature matrix $\mathcal{X}$, where $d$ is the number of features defining the semantic node. The GCN produces a node-level $n\times f$ output feature matrix $\mathcal{H}$, where $f$ is the number of output features per node. 

In our pipeline, the GCN takes an $n\times16$ input feature matrix and produces an $n\times18$ output feature matrix, called \textit{semantic hints} $\mathcal{H}$. The input features include: the one-hot-encoded labels of the semantic segments (with 9 semantic classes, see \ref{input_feat}), median and standard deviation for each R, G and B channel, and the median luminance value, all computed for the pixels belonging to the corresponding semantic segment.

Each layer $l$ of the GCN can be represented as a function:
\begin{equation}
    Y^{(l+1)} = f(Y^{(l)}, \mathcal{E}) = \sigma\left( \mathcal{E}Y^{(l)}W^{(l)}\right)\,,
\end{equation}
where $Y^0 = \mathcal{X}$, $Y^{(L)} = \mathcal{H}$, and $L$ is the last layer. $\mathcal{E}$ is the edge representation in form of an adjacency matrix, $W^l$ is the weight matrix of layer $l$ of the GCN and $\sigma(\cdot)$ is a non-linear activation function which, in our case, is Leaky-ReLU.

\begin{figure*}[t]
\centering
  \includegraphics[width=.74\textwidth]{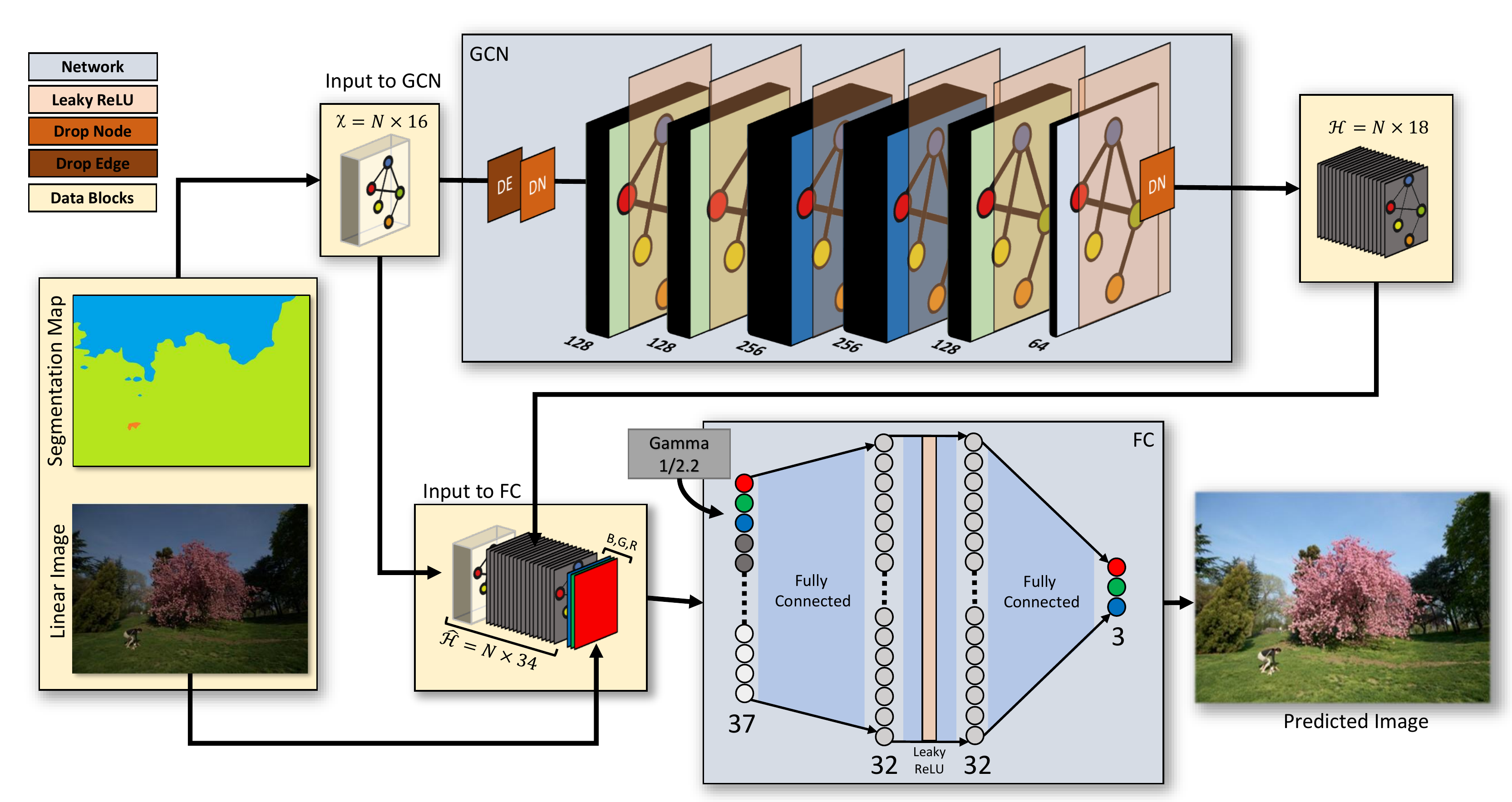}
  \caption{\textit{G-SemTMO details}: G-SemTMO has 4 data blocks and 2 network blocks. Using the input linear image and segmentation map we obtain a connected graph of $N$ semantic nodes and an input feature matrix $\mathcal{X}$. $\mathcal{X}$ and the node adjacency matrix is forwarded to the first network block GCN. The GCN has 6 graph convolutional layers followed by an activation layer of Leaky-ReLU. \textit{DropEdge}\cite{rong2019dropedge} and Node dropouts are used to prevent over-fitting. The GCN outputs latent semantic hints $\mathcal{H}$ with 18 hints per node. Broadcasted features $\mathcal{X}$ and $\mathcal{H}$ stacked together ($\hat{\mathcal{H}}$) and the input linear RGB create the final data block which is forwarded to the final network block FC. The FC has 2 fully connected layers with an activation Leaky-ReLU layer between the two. A gamma curve of 2.2 is applied to the input of the FC and the output is the tone mapped image.}
  \label{arch}
\end{figure*}

\subsection{Tone Mapping Module}
The tone mapping module constitutes of a Fully Connected (FC) network, which acts as a 3D lookup table to map each input linear RGB pixel to the output display-encoded RGB pixel. Supplemental inputs allow this function to be local and semantics aware: the contextual information in form of $n\times18$ semantic hints $\mathcal{H}$ from the GCN is passed in addition to the spatial information from the $n\times16$ input feature matrix $\mathcal{X}$. The combined semantic information $\hat{\mathcal{H}}$ from the resulting $n\times34$ matrix is spatially arranged with the input linear image such that each pixel in the image corresponds to $37$ values: the 3 RGB channels and a 34-element semantic hint-feature vector. 
Consequently the FC trains over this 37 channel data to learn a mapping function:
\begin{equation}
    f(I_R,I_G,I_B,\hat{h_1}, \hat{h_2} ... \hat{h}_{33},\hat{h}_{34}) = O\,.
\end{equation}
We train to minimise the $L_1$ difference in pixel values for all pixel positions $\{i,j\}$ and colour channels between the predicted, $O$, and reference, $R$, images: 
\begin{equation}
    \mathcal{L} = \sum_{i,j}\sum_{c \in \{R,G,B\}}
    \Big\lvert R_{c,i,j} - O_{c,i,j} \Big\rvert\,,
    \label{loss}
\end{equation}
where both $R$ and $O$ are gamma-encoded RGB images in ITU-Rec.709 color space.

\subsection{The Implementation Details}
\subsubsection{Preparing the image dataset}
MIT-Adobe FiveK dataset \cite{fivek} provides a set of 5000 high resolution RAW images and their manually retouched versions provided by 5 expert photographers (A, B, C, D, E). Prior work on image enhancement uses retouched versions created by expert C \cite{hwang2012context, yan2016automatic, fivek}. Gharbi et al.~\cite{gharbi2017deep} use all 5 expert versions for their HDRNET but mention the inconsistencies among the expert retouches. They mention that expert B is more self-consistent and easier to learn for the network. 

We initially choose expert E based on our subjective aesthetic preference of retouched results. However, we show in \secref{other_experts} that our architecture can learn irrespective of the choice of expert photographer. We observe that the dataset contains a significant number of images with large portion of saturated pixels in the RAW images. The RAW conversion software (Adobe Lightroom) reconstructed those pixels to non-unique colors in the retouched images. As such saturated pixels may lead to inconsistent learning, we filter images with high number of saturated pixels before training. We empirically set a threshold of $3\%$ pixels per image to filter pixels where any of the RGB channels values are above a normalised tonal value of $.99$. This provides us with $4205$ 16-bit linear color images and their retouched versions for our training. We use the `as-shot' white balance applied by the camera while exporting the linear images. For training, we resize images to the resolution of $100\times100$ pixels.

\subsubsection{Preparing the input features}
\label{input_feat}
The next step is to generate input feature space for each image-graph representation. Global attributes and overall visual cues such as the average luminance or standard deviation of intensity values could inform decision on image enhancement. Based on this idea, Yan et al.~\cite{yan2016automatic} use both global and contextual feature descriptors for their image enhancement. We use similar attributes corresponding to each semantic region of the image. First, we use FastFCN semantic classifier \cite{wu2019fastfcn} pre-trained over ADE20K dataset \cite{ade20k} to generate segmentation maps. ADE20K provides a dataset with $150$ annotated labels which results in a very fine-grained semantic breakdown of an image. However, we realise that, in the use-case of digital photography, the semantic abstraction which drives decision on image edits is not as fine-grained. Therefore, we merge the fine labels to a coarser semantic abstraction based on the work of Goswami et al.~\cite{goswami2020tone}. The 9 coarse labels: \textit{sky, mountain (terrain), vegetation, water, human subject, non-living subject, city, indoor-room, others} fit the use-case of digital photography better. The segmentation maps are generated at full resolution and consequently resized to match the training image resolution of $100\time100$. The spatial arrangement of the segments are stored in the edge descriptor $\mathcal{E}$ in pytorch coordinate format (COO) for the GCN. Furthermore, we compute  attributes for each segmented region: the median and standard deviation of RGB values, the median luminance and the 9-class one-hot encoded semantic labels for each semantic node.

\subsubsection{GCN and semantic hint generation}
Fig.~\ref{arch} presents our architecture in detail. The GCN based Semantic Hint module has 6 graph convolutional layers generating $128, 128, 256, 256, 128$ and $64$ latent features respectively. Each convolutional layer is followed by an activation function of Leaky-ReLU. To prevent overfitting the model, we drop nodes in form of dropout layers before the first convolutional layer and after the last convolutional layer with probability of $0.2\text{ and }0.5$ respectively. Furthermore, we apply a DropEdge~\cite{rong2019dropedge} with a probability of 0.2 before the first dropout.

\subsubsection{Prediction using FC}
The FC acts as a lookup table which maps 37 input values to 3 output RGB values. During training the input to the FC is a 2D array $10000\times37$ containing all pixels in the image and their corresponding hints. We observe that applying a power of $1/2.2$ to the input of the FC helps it learn the non-linear mapping better. The FC has two fully connected hidden layers with 32 neurons each separated by a layer of Leaky-ReLU activation function. The output of the FC is the predicted non-linear RGB value. Due to the design of pixel prediction, the inference can be obtained on a high resolution image instead of $100\times100$.

\subsubsection{Blending}
The predicted output RGB values show visible inconsistencies at the border of semantic regions due to 1) the difference in tone mapping function across regions and 2) lack of smooth transition and segmentation precision of the FastFCN algorithm. In order to incorporate pixel precision, we utilise a shared alpha matting technique \cite{GastalOliveira2010SharedMatting} and draw inspiration from the semantic framework idea of Goswami et al.~\cite{goswami2020tone} which involves stacking normalised fuzzy segmentation maps of each semantic region and blending the tonal modification.

To create the framework, we first breakdown a segmentation map containing \textit{n} unique labels into \textit{n} binary maps. Shared matting \cite{GastalOliveira2010SharedMatting} converts each binary map to a fuzzy alpha map using a trimap obtained by dilation of the segment in the binary map with a disk structuring element of the radius of 25 pixels. 
Each alpha map is processed by a bilateral filter (we set pixel neighborhood diameter d=$50$ and color parameter $\sigma=30$) to remove discontinuities if introduced due to the morphological operations in matting. The alpha maps are stacked along the \textit{z}-axis and normalised to complete the semantic framework \textit{(S)}. The FC is used to infer \textit{n} images, one for each semantic hint where the same hint is used for all pixels. Stacking the \textit{n} images similarly provides an image framework \textit{(F)}.
The weighted summation of the two frameworks provides us the blended image result.
\begin{equation}
    O_{blended} = \sum_i^n S_i\cdot F_i 
\end{equation}

\subsubsection{Training procedure}
\label{sec:training_details}
We use 4000 resized images out of the selected 4205 to train our network and keep 106 images for validation and 99 for inference. The weights and biases are optimized by minimising the loss defined in Eq.~\ref{loss}. The weights are further regularised with a weight decay of $5e-4$. We optimize the network parameters using ADAMW solver \cite{adamw}. We train in batch size of 1 due to the variable structure of the graphs and the learning rate is scheduled to vary with the epoch. We train for 250 epochs with a learning rate of $10^{-3}$ between epoch 0 and 75, of $10^{-4}$ between epoch 75 and 150, and $10^{-5}$ from 150 onwards. We implement our architecture using PyTorch \cite{NEURIPS2019_9015} and PyTorch Geometric \cite{Fey/Lenssen/2019} on an Nvidia RTX2060 GPU. The training takes about 24 hours. 

\section{Ablation Study}
\label{ablation_desc}

To analyze the importance of each component of our method, we conduct two ablation experiments in addition to the proposed G-SemTMO. We observe in literature that tone mapping approaches work better than existing methods when explicit semantic information is provided as input \cite{goswami2020tone}. Our hypothesis is that it can be improved further when contextual semantic information is supplied in conjunction to the learning pipeline. We designed our ablation experiments to incrementally modify the sophistication of semantic information introduced to the learning pipeline as follows:\\
\textit{\textbf{Ablation 1: 3D LUT-Global mapping} Fully connected neural network (FC) without any semantic information.}\\ 
We utilise the FC architecture of our Tone Mapping Module to learn the mapping from linear RAW images to expert retouched images. No additional semantic information is provided and GCN is not used.\\
\textit{\textbf{Ablation 2: 3D LUT-Local Semantic mapping} FC with semantic information.}\\ 
We train the image pairs over the FC architecture similar to Ablation~1. But for every pixel semantic information is added. A vector of size $19$ is provided -- $3$ colour channel values and semantic-specific input features of size $16$ similar to the input to GCN (refer to \secref{semantic_hint_module}). GCN is not used.
%We train the image pairs over the FC architecture as in Ablation~1 but supply the 37-element vector with semantic hints to the FC network. No GCN is used in Ablation~2.
%. However, in this case as an input to the FC, we introduce spatial semantic information identical to the input to our GCN architecture $\mathcal{X}$.
\\
\textit{\textbf{Ablation 3: Graph based Semantic mapping} G-SemTMO.} \\
This ablation consists of the full architecture with GCN, as explained in \secref{pipeline desc}. GCN uses semantic-specific input features to provide semantic hints to FC.
%We introduce graph based learning to provide the contextual information in addition to the spatial semantic information for the training pipeline in form of semantic hints. The pipeline is the same as our method described in \secref{pipeline desc}. It is important to remind the difference between Ablation~2 and Ablation~3. In the latter case, the GCN and the input to the GCN is also passed to the FC along with the semantic hints.

For all the ablation studies the models are trained using the same hyper-parameters. They are trained on 3000 training image pairs and validated on 20 image pairs for 250 epochs with a learning rate of $10^{-4}$ and a weight decay of $5e-4$. To report test results, we compute the mean pixel HyAB perceptual colour distance \cite{abasi2020distance} and the PSNR for the prediction results of 99 test images. We used HyAB rather than CIE~DeltaE as it was shown to better capture luminance differences. 

\begin{figure}[]
\centering
  \includegraphics[width=.68\linewidth]{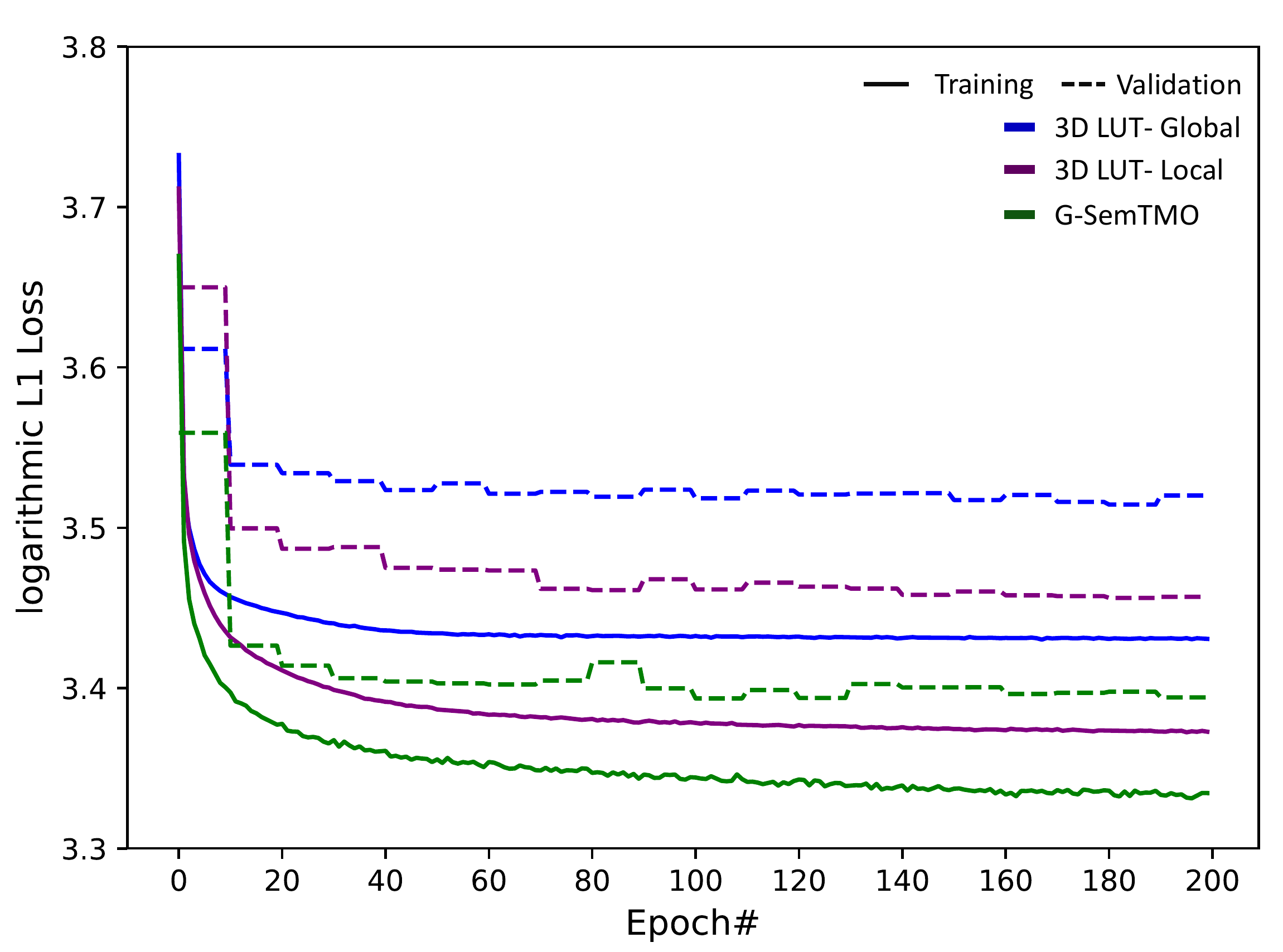}
  \caption{The convergence plots for the ablation study.}
  \label{fig:ablation training}
\end{figure}

\begin{figure*}[]
    \footnotesize
    {\centering
    \begin{tabular*}{\textwidth}{cccc}
    % \toprule
        3D LUT Global mapping & 3D LUT Local mapping & G-SemTMO & Ground Truth \\
    % \midrule
      \includegraphics[width=.225\linewidth]{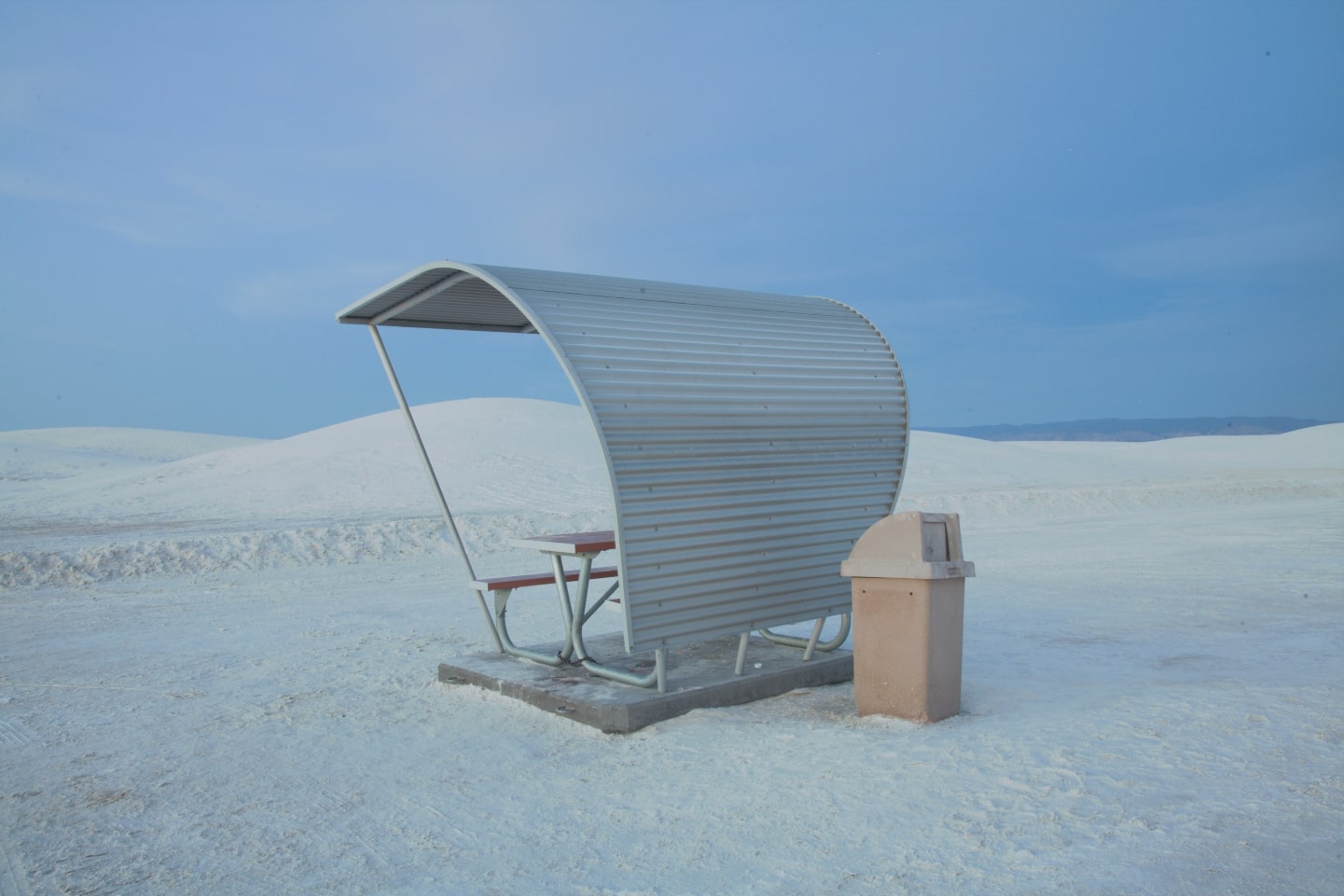} & \includegraphics[width=.225\linewidth]{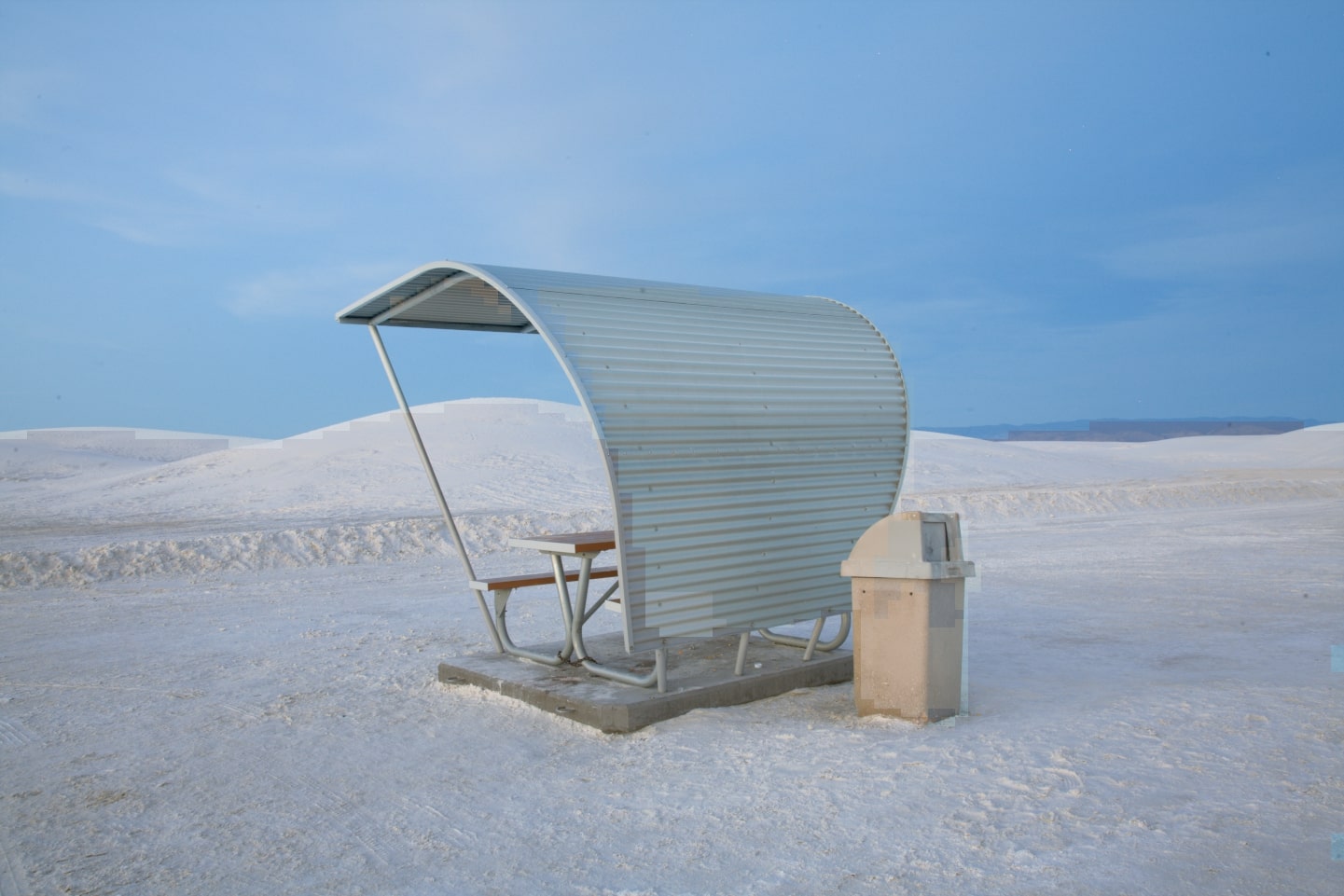} & \includegraphics[width=.225\linewidth]{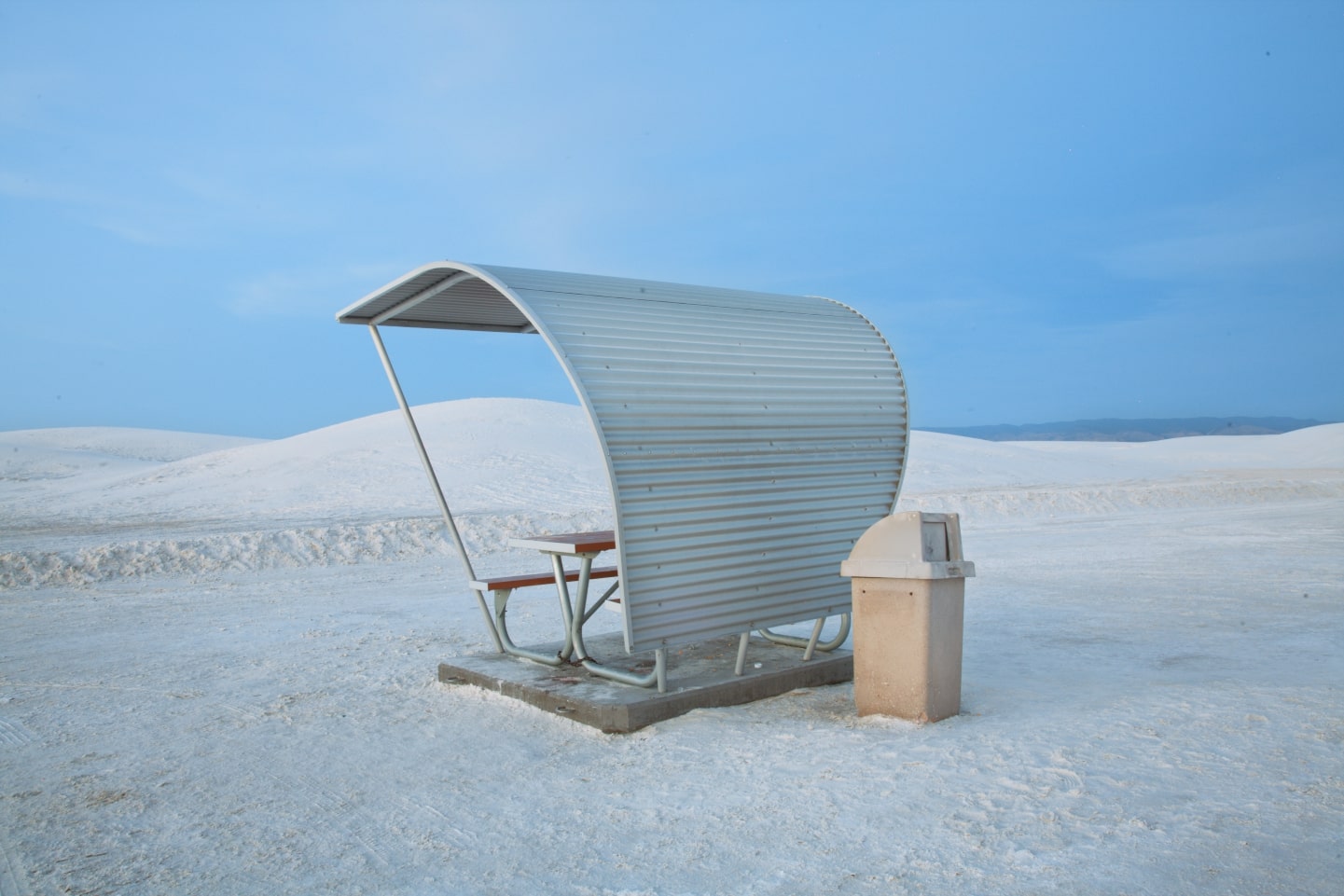} & \includegraphics[width=.225\linewidth]{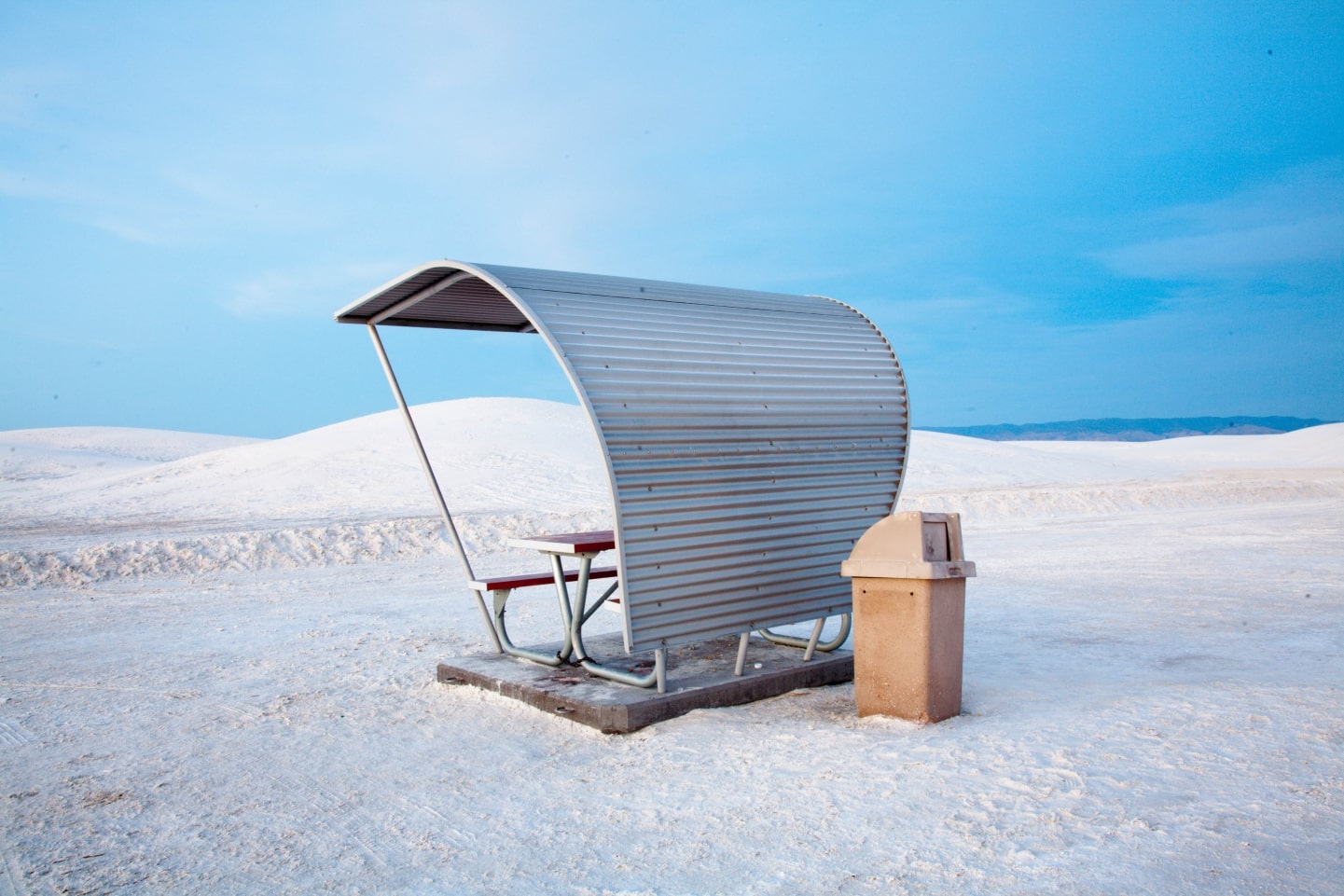} \\ 
      HyAB: 12.12 \enspace PSNR: 16.95 & HyAB: 10.55 \enspace PSNR: 18.21 & \textbf{HyAB: 5.85 \enspace PSNR: 21.91}& \\
      \includegraphics[width=.225\linewidth]{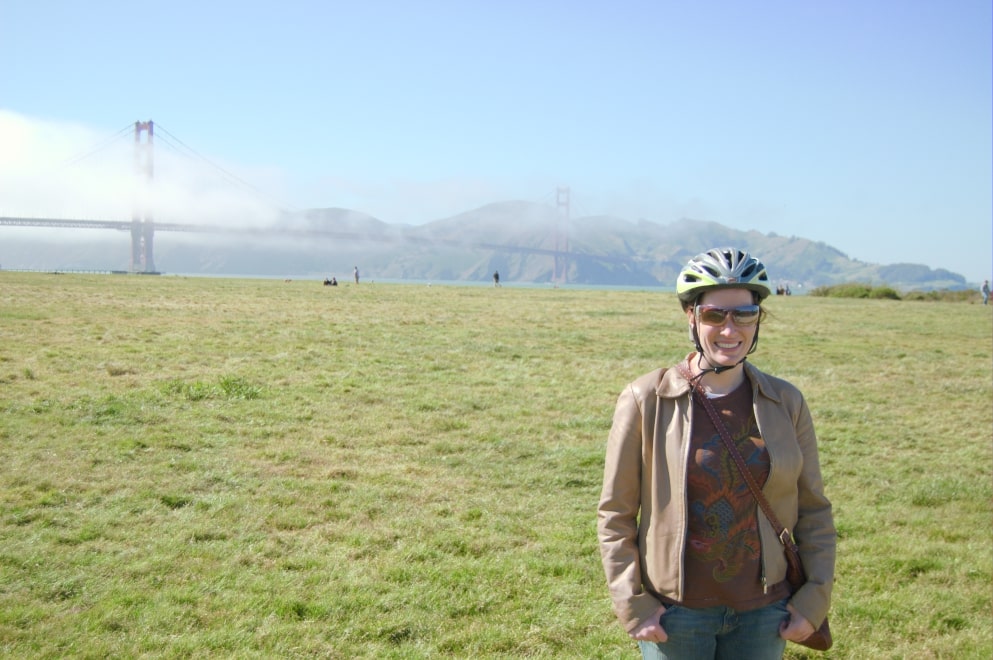} & \includegraphics[width=.225\linewidth]{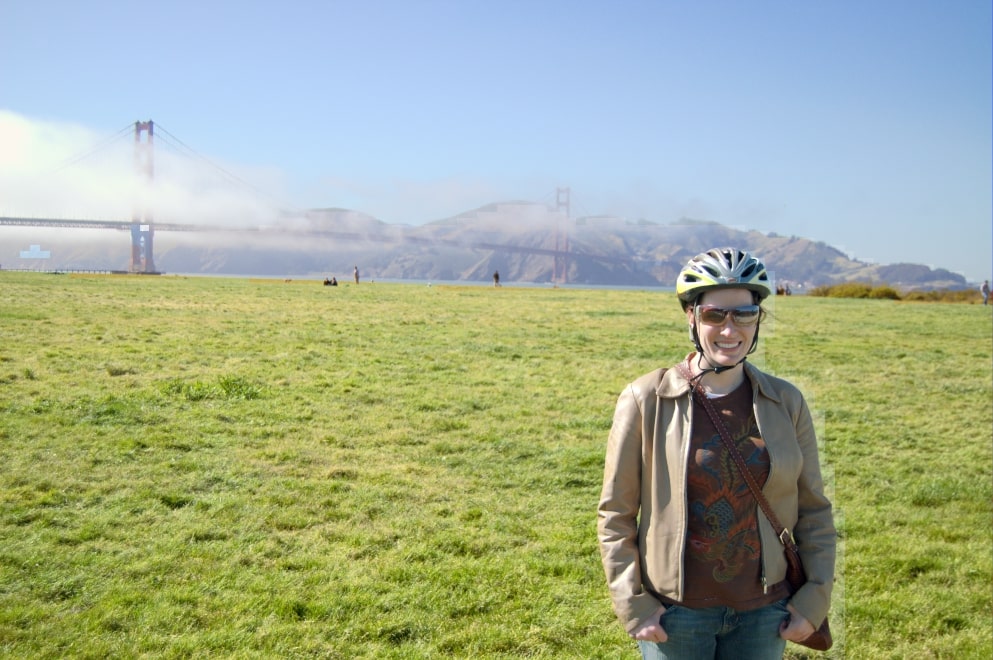} & \includegraphics[width=.225\linewidth]{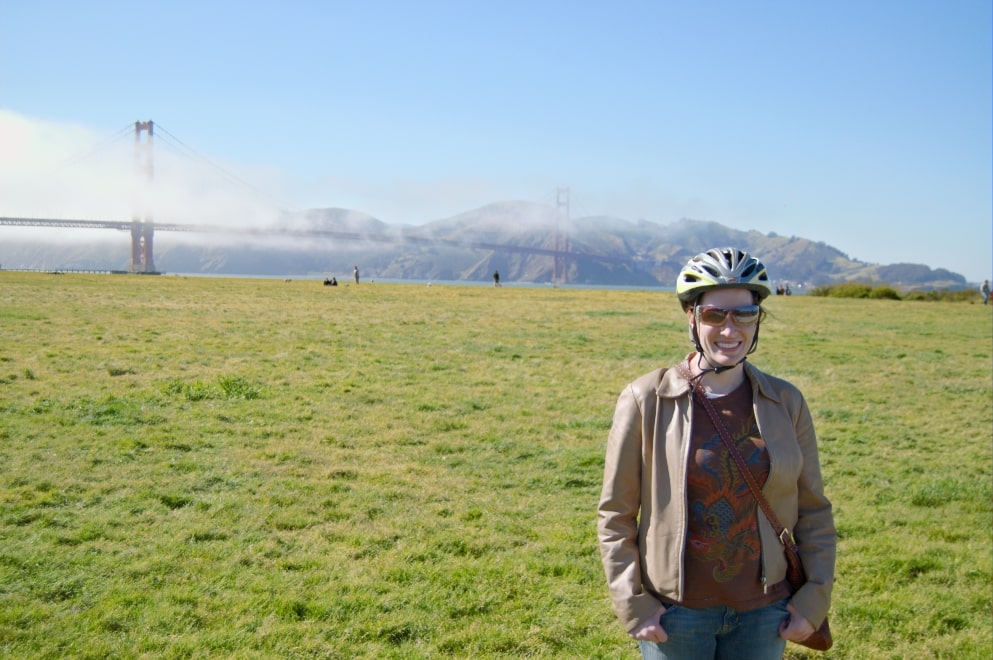} & \includegraphics[width=.225\linewidth]{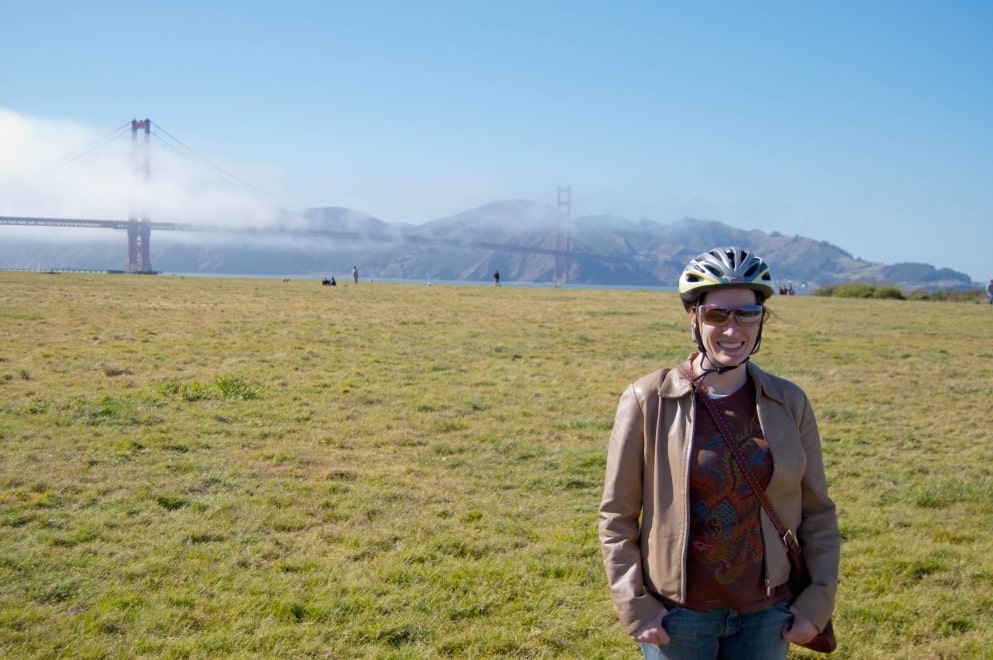} \\ 
      HyAB: 9.66 \enspace PSNR: 19.69 & HyAB: 6.08 \enspace PSNR: 23.33 & \textbf{HyAB: 3.59 \enspace PSNR: 24.53}& \\
      \includegraphics[width=.225\linewidth]{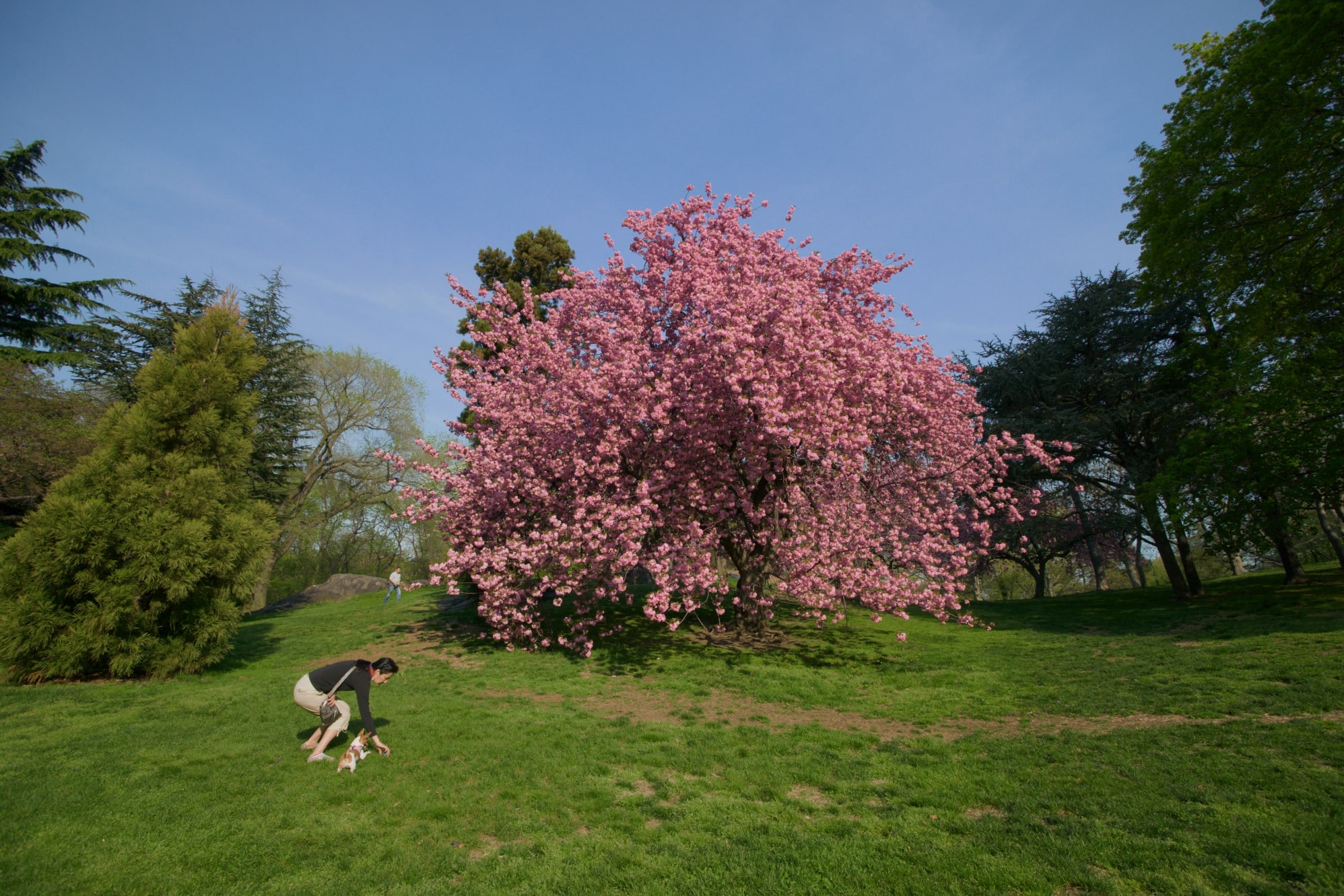} & \includegraphics[width=.225\linewidth]{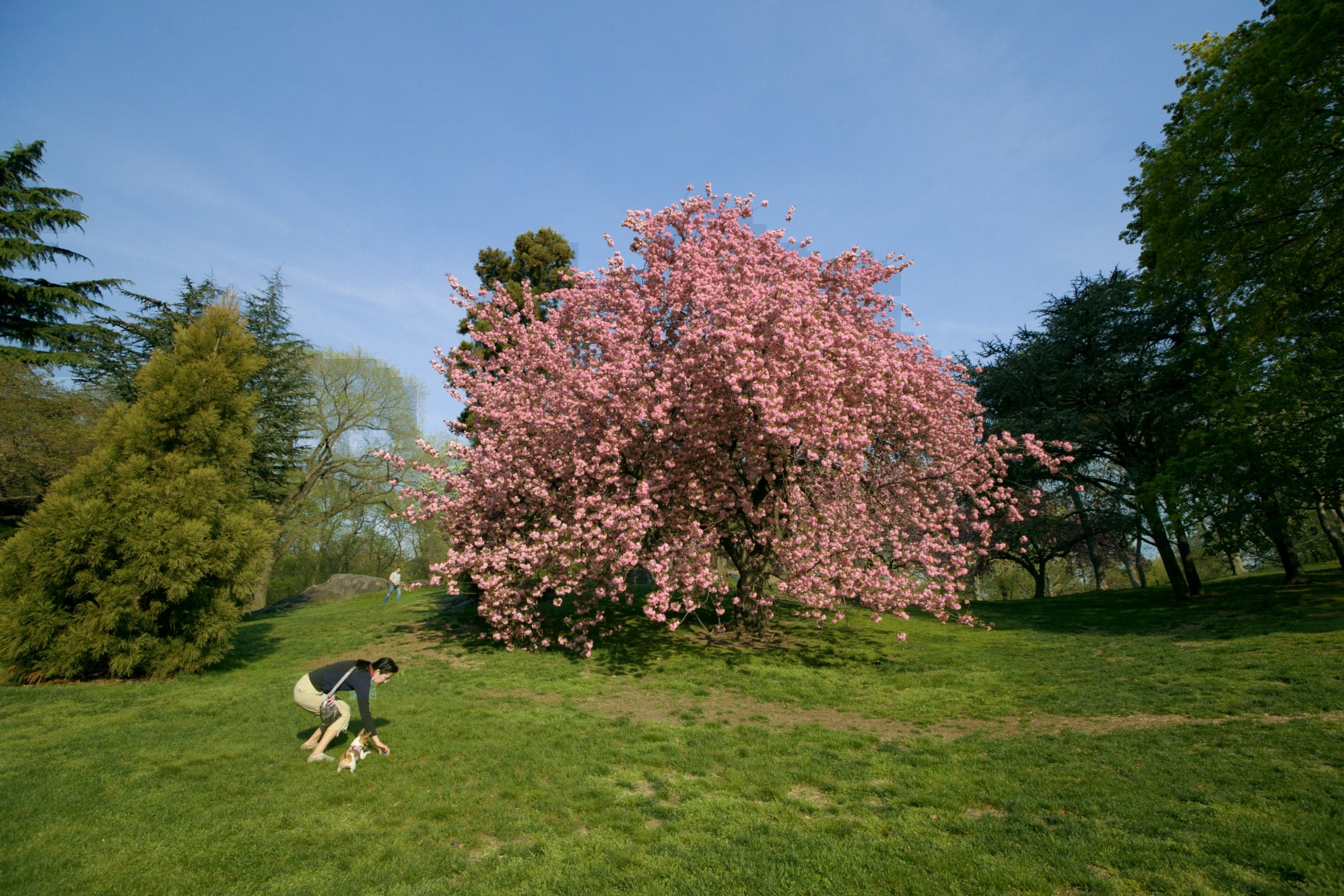} & \includegraphics[width=.225\linewidth]{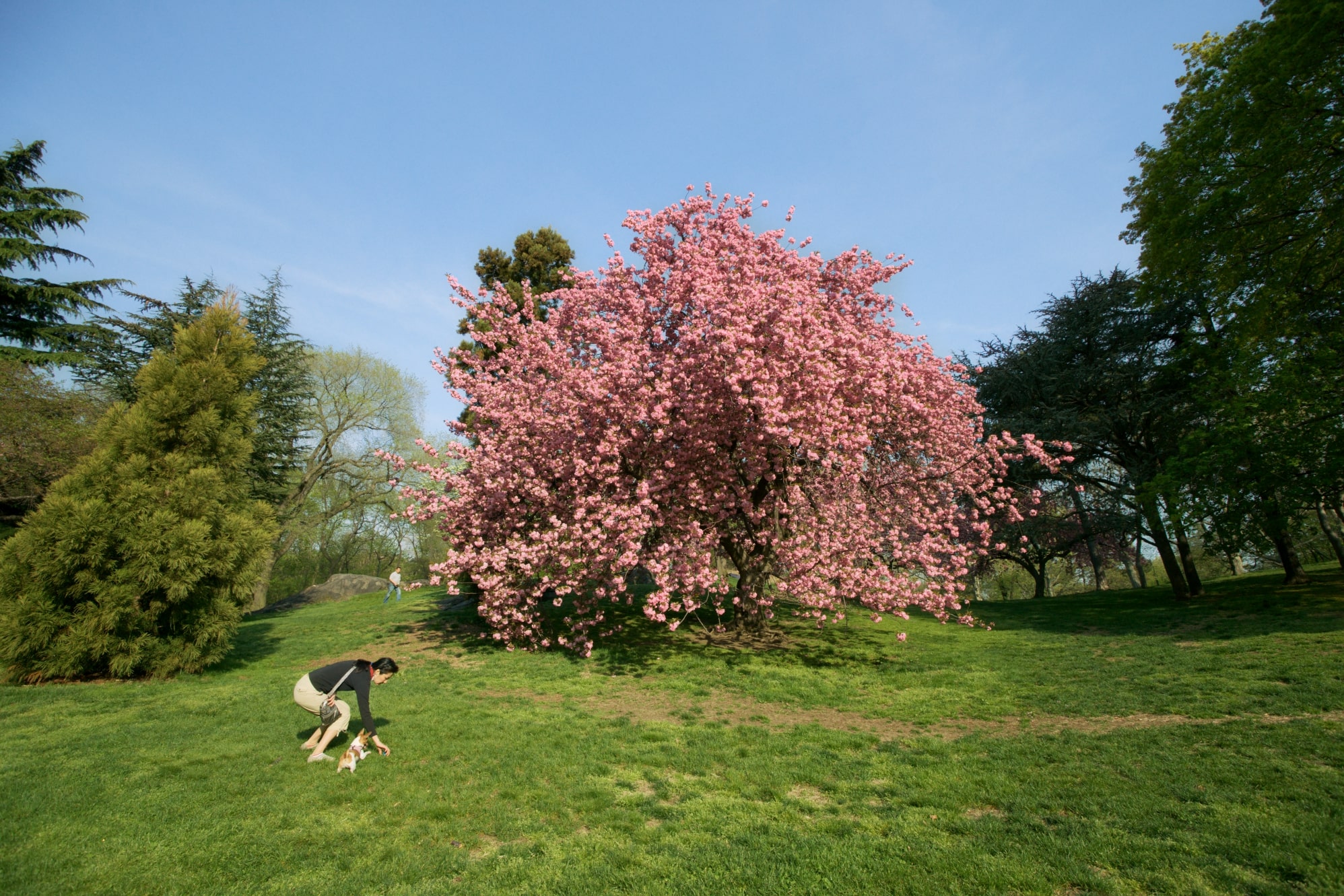} & \includegraphics[width=.225\linewidth]{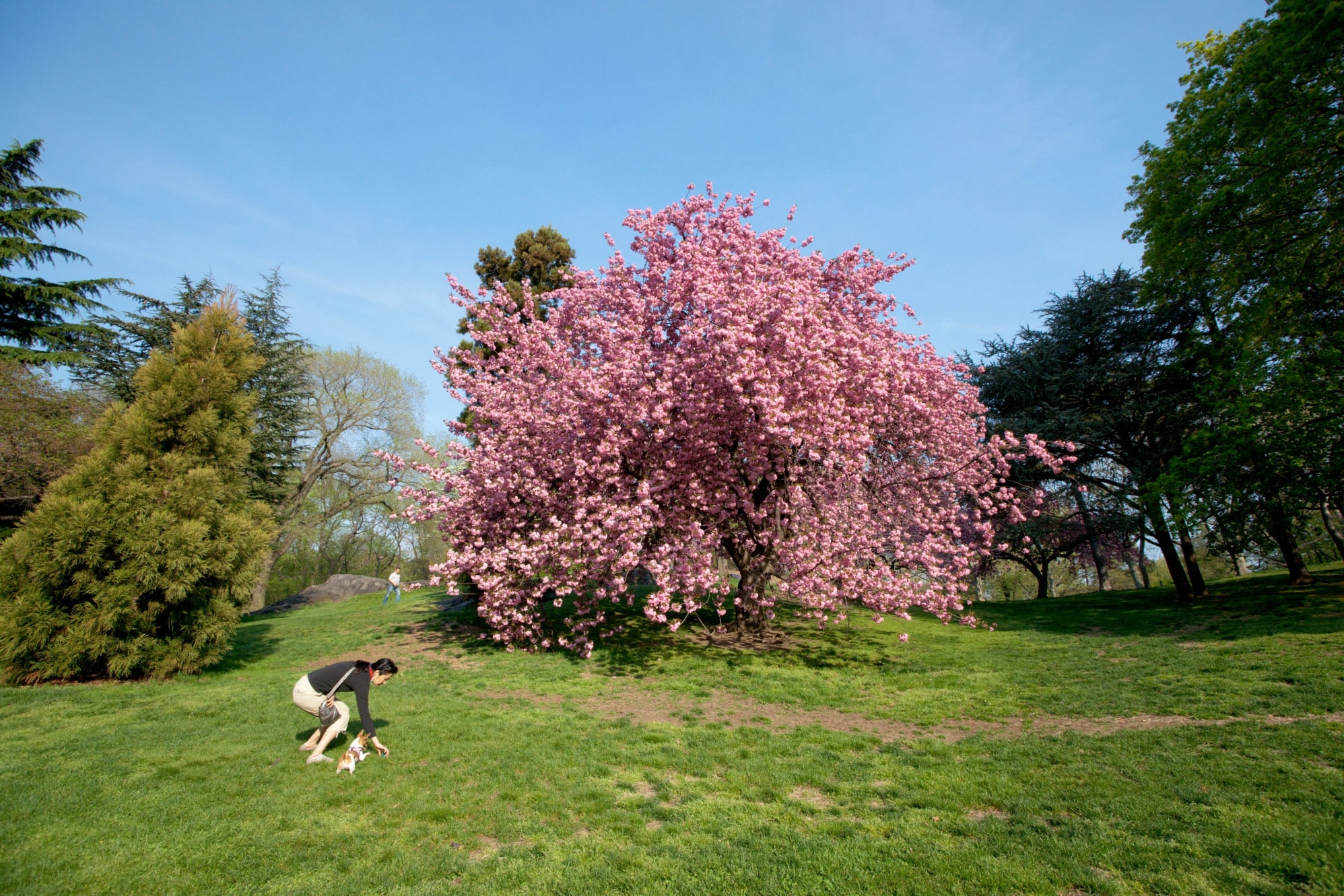} \\ 
      HyAB: 10.92 \enspace PSNR: 18.01 & HyAB: 8.00 \enspace PSNR: 19.79 & \textbf{HyAB: 5.29 \enspace PSNR: 21.27} & \\
      
    % \bottomrule 
    \end{tabular*}}
    \caption{\textit{Ablation comparisons.} We present tone mapped results from the 3 ablation studies for 3 images from the FiveK dataset: \textit{a4886, a4986} and \textit{a5000}. \textit{Left to right}: 1) \textit{Ablation 1} 3D LUT Global tone mapping. 2) \textit{Ablation 2} 3D LUT Local tone mapping with semantic-specific information. 3) Proposed G-SemTMO which considers semantic information as well as the contextual information from spatial arrangement of semantic labels using graph convolutions. 4) The manually retouched version of the image produces by expert E \cite{fivek}. The HyAB and PSNR objective metric scores for each tone mapped image validates the advantage of graph-based learning over the other ablation studies.}
    \label{fig:Ablation_table}
\end{figure*}

\textit{Observations:} \figref{ablation training} illustrates the training and validation loss curves across the three studies. The curves confirm our hypothesis that enriching the feature space with contextual semantic information improves the performance of the model. Across the three ablation studies, we observe that the model with the the full semantic information results in lower training and validation loss.
\figref{Ablation_table} presents 3 images from the FiveK \cite{fivek} dataset tone mapped by the networks from the ablation study and their respective HyAB color distance and PSNR scores. Based on subjective assessment, we conclude that our proposed graph-based learning produces results much closer to the ground truth for the selected images. 

\begin{figure}[]
\centering
  \includegraphics[width=.485\linewidth]{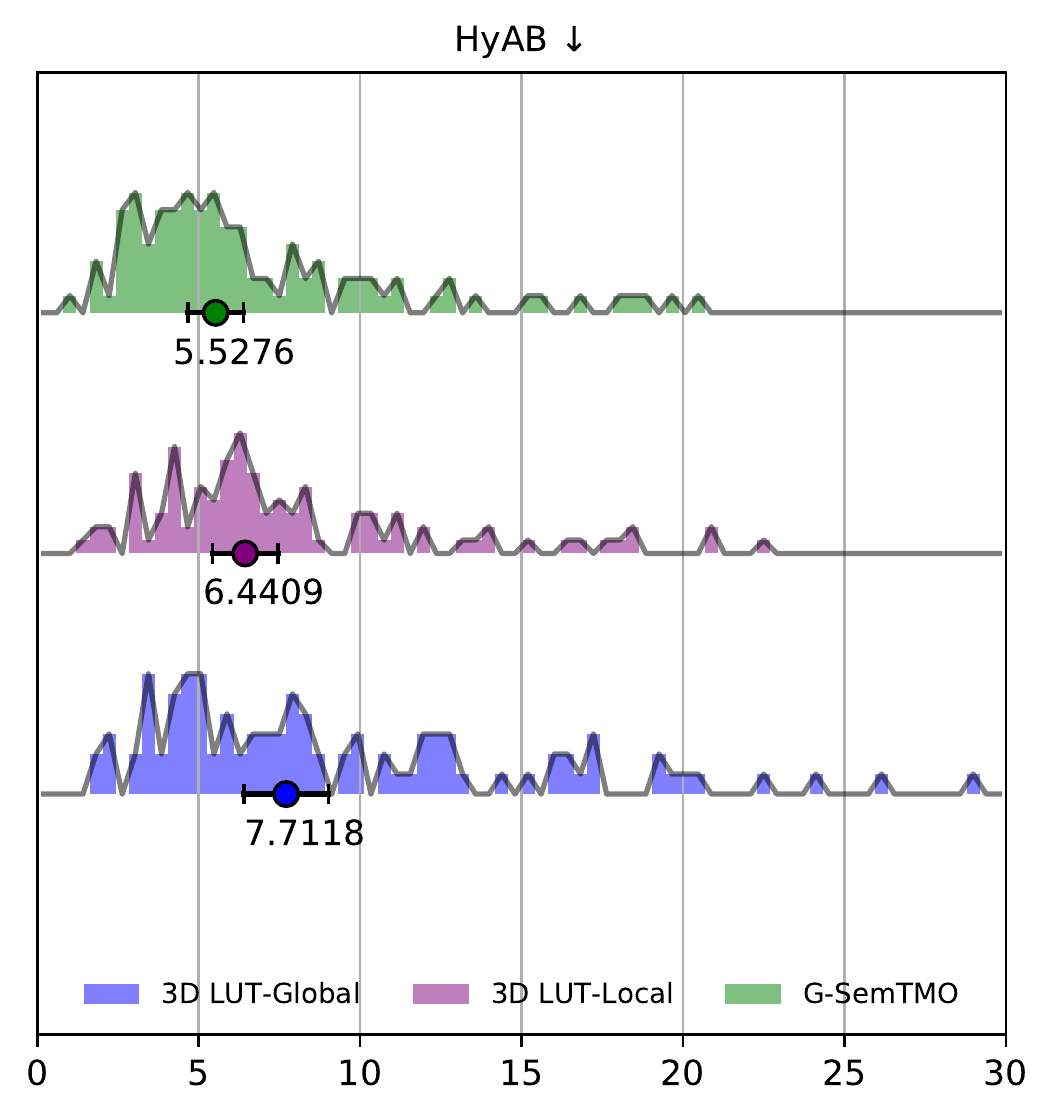}
  \includegraphics[width=.485\linewidth]{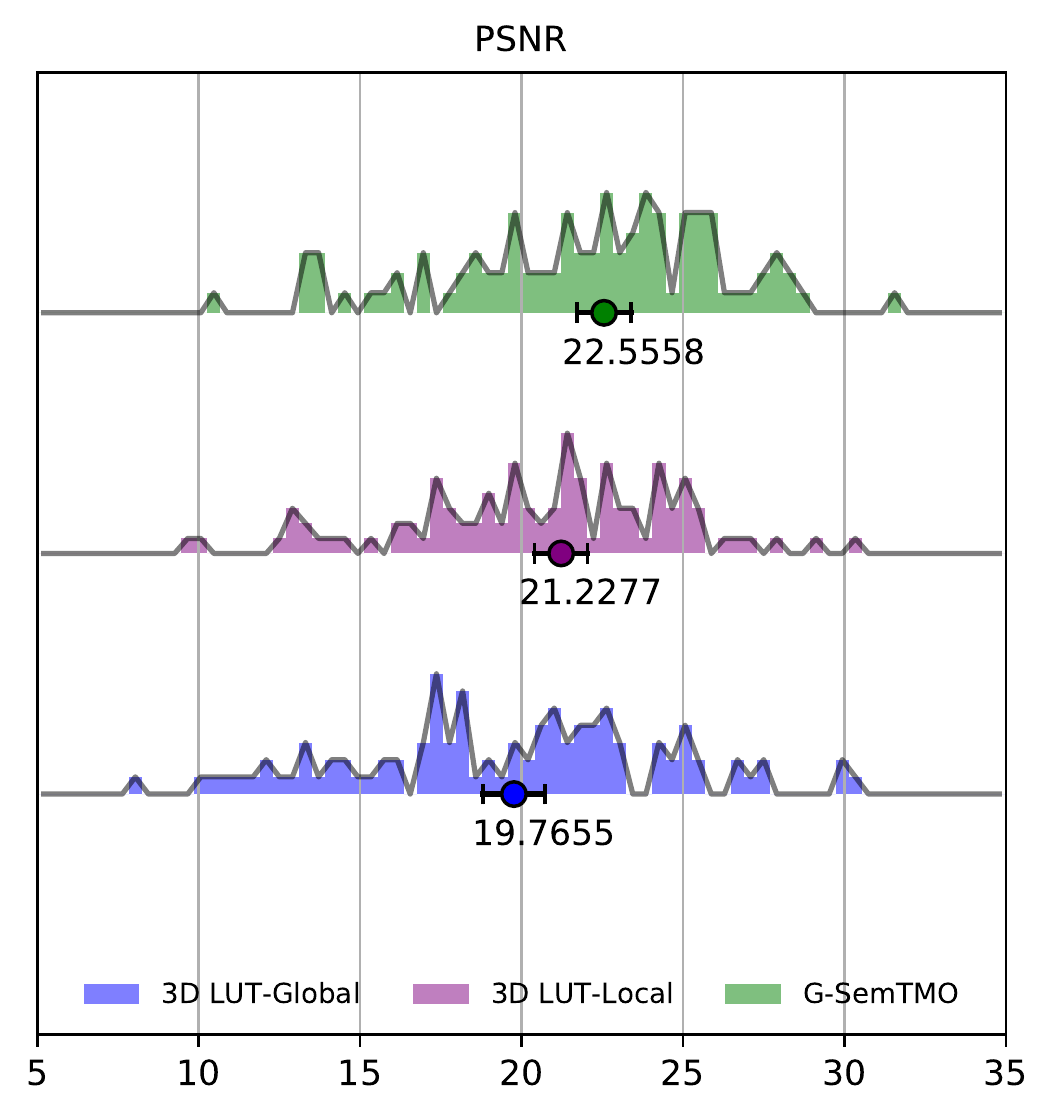}
  \caption{The histograms of HyAB and PSNR scores for the 3 ablation studies are presented. The histograms correspond to score distribution over 99 test images. The median of the distribution is plotted with a solid circle with a confidence interval of 95\%.}
  \label{fig:ablation metric}
\end{figure}

\figref{ablation metric} plots the histogram of HyAB and PSNR objective scores across 99 test images for each ablation study. Additionally, we plot the median for each histogram with its confidence interval of $95\%$. We observe that the proposed G-SemTMO prediction gets closest to the images retouched by expert E with a median perceptual colour distance score of $5.53$. It also receives better PSNR evaluation than the other two ablations. 

\begin{figure*}
  \includegraphics[width=\linewidth]{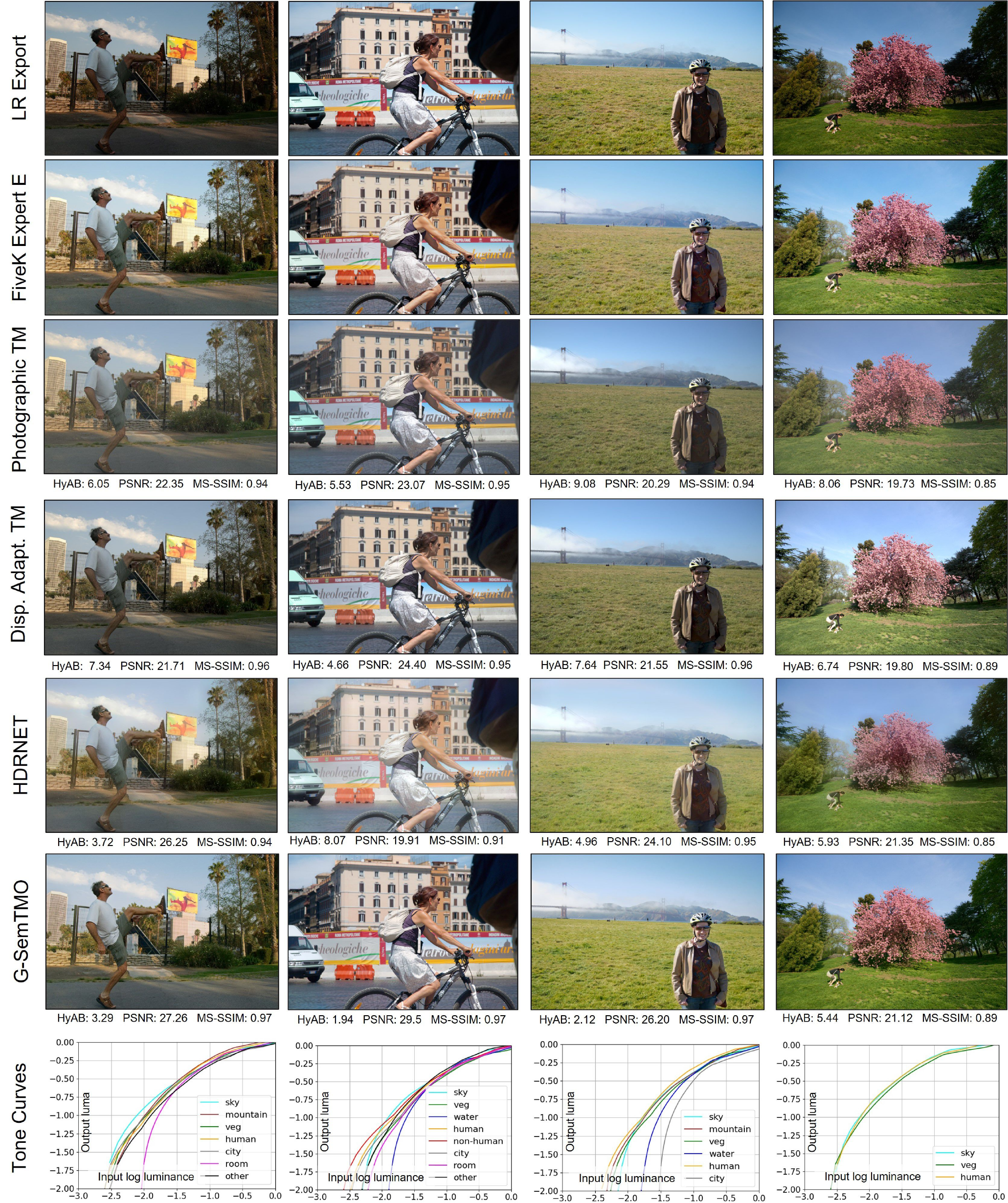}
  \caption{\textit{Left to Right}: 4 selected images from FiveK dataset. \textit{Row 1}: Raw images exported using Lightroom without modification. \textit{Row 2}: Target images modified by expert E \cite{fivek}. \textit{Row 3-6}: Selected TMOs with HyAB, PSNR and MS-SSIM scores. \textit{Row 7}: Tone curve applied per segment by G-SemTMO.}
\label{fig:comparison}
\end{figure*}

\section{Learning Global Image Enhancements}

\subsection{References for Comparison}
In this section, we present images tone mapped using G-SemTMO and compare the results against the prediction of another machine-learning-based method, HDRNET \cite{gharbi2017deep}, retrained on the same images as our method. We also include the results of 4 traditional TMOs: Photoreceptor TM \cite{reinhard2005dynamic}, Photographic TM \cite{reinhard2002photographic}, Display Adaptive TM \cite{mantiuk2008display} and Bilateral TM \cite{durand2002fast}. The traditional TMOs do not allow for training and they are included in our comparison to show the difference between trained and non-trained tone mapping. We present our observation based on our subjective assessment and validate them using objective metrics. 

Since we are unable to train the official HDRNET Tensorflow implementation due to rather old version of the dependencies, we rely on the PyTorch re-implementation by Jinchen Ge~\cite{jinchen}. Gharbi et al.~\cite{fivek} use FiveK dataset to learn style transfer and their network was trained using image pairs comprising of 8-bit input images without corrections and 8-bit images retouched by experts. However, as per author suggestions, we use their network architecture to train for end-to-end tone mapping using 16-bit linear images as input and 8-bit retouched images as output. To generate the results for the 4 classical TMOs, we used pfstools\footnote{http://pfstools.sourceforge.net/pfstmo.html.} software.

To assess the tone mapping results, we use 4 objective metrics: PSNR, the hybrid perceptual colour distance metric HyAB \cite{abasi2020distance}, Mutli-scale Structural Similarity Index (MS-SSIM) \cite{wang2003multiscale} and the Visual Difference Predictor for HDR images HDR-VDP-3 \cite{mantiuk2011hdr}.
% \RM{The short description of the metrics could be appreciated, but the one below is rather long. PSNR needs to introduction. }
More precisely, we choose HyAB, a perceptual color distance metric, to assess the closeness of color reproduction to the ground truth. It uses \textit{$L_1$} norm of L* and \textit{$L_2$} norm of a*b* in the CIELAB colour space to measure color distance from the reference image. It has shown good agreement to subjective preference for small colour deviations. A smaller HyAB score suggests better quality.
To evaluate the reproduction of structural details and local contrast preservation, we use MS-SSIM, which is a multiscale version of traditional structural similarity index. A higher MS-SSIM score suggests a higher measure of structural similarity resulting in better perceptual quality.

Furthermore, for overall quality, we choose traditional PSNR and the HDR-VDP-3 (v3.0.6)\footnote{https://sourceforge.net/projects/hdrvdp/files/hdrvdp/.} Quality correlate (Q) score. It is a measure of the magnitude of distortion corresponding to visibility rather than the mathematical distance between the pixels. The HDR-VDP-3 score attains a maximum of 10 for best perceptual quality and gets lower for poorer reconstruction. 

\subsection{Observations}
\figref{comparison} presents the results for 4 images from the MIT Adobe FiveK dataset \cite{fivek} (from the testing set). In the top row, we reproduce the original RAW images exported via Lightroom without any modification except for the standard gamma encoding. The second row contains the images manually retouched by \textit{expert E} (used for training HDRNET and G-SemTMO) and the following rows contain the results of each operator. Objective metric results, using HyAB, PSNR and MS-SSIM with respect to Expert E, are also indicated. The last row contains the plots of the per segment gray-scale tonecurves, produced by G-SemTMO for each semantic region. The tone curves are generated by mapping input grayscale values (where $R=G=B$) to output color and then computing the luma value. Our first observation is that, for the selected images, G-SemTMO produces results closer to the expert retouched images than HDRNET trained on the same data. The comparison with traditional operators is for illustrative purposes only as they have never been trained and are not meant to reproduce the results of expert~E. 

\begin{figure}[t]
\centering
  \includegraphics[width=.76\linewidth]{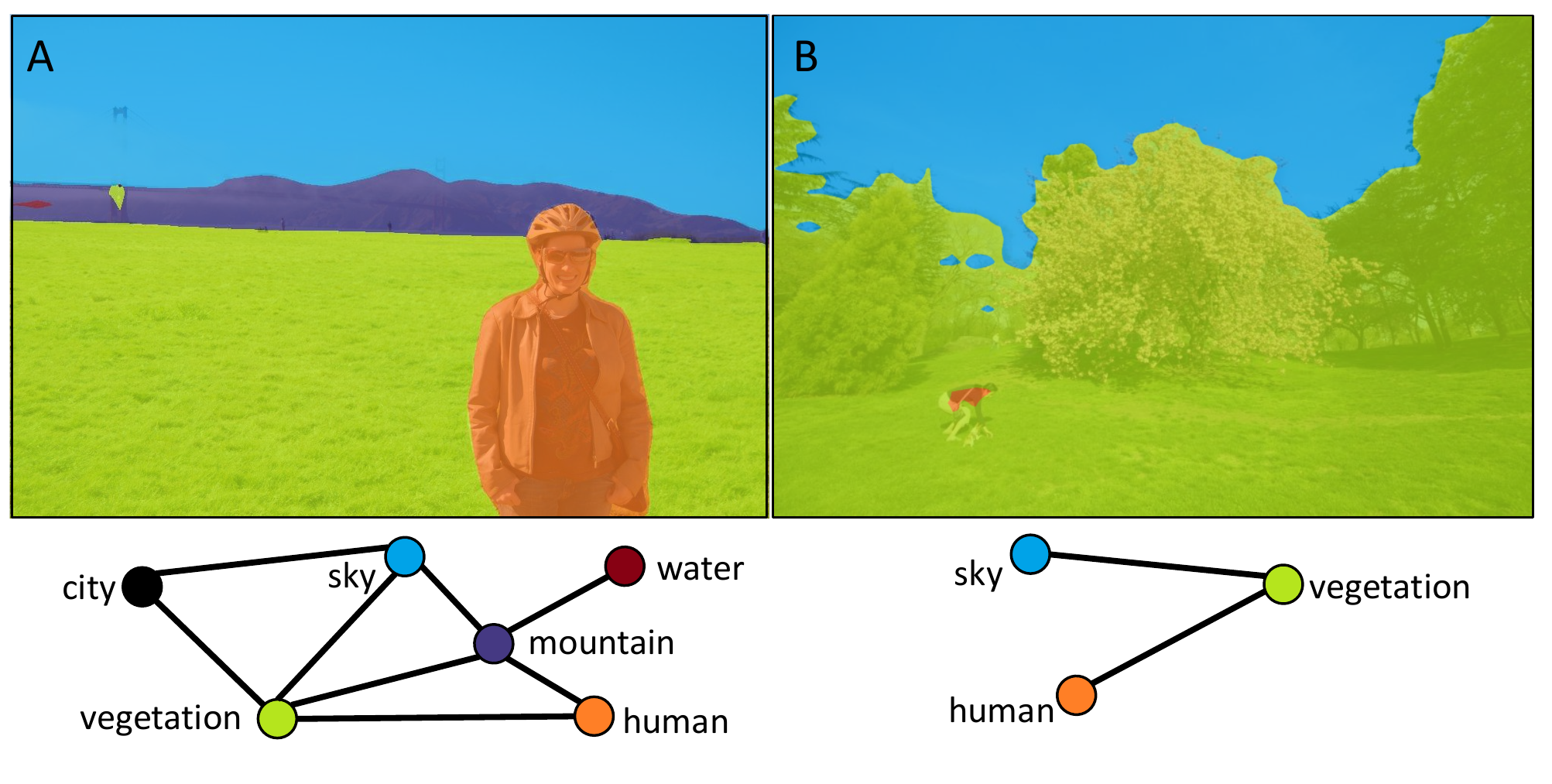}
  \caption{Neighborhood based tonal adjustment. \textit{Left to Right:} Image A (a4986), Image B (a5000) and their corresponding graph representation of semantic labels. The neighborhood of \textit{vegetation} is different in A from B, the predicted tone curve would be different too.}
  \label{fig:tc_explain}
\end{figure}
\begin{figure*}[b]
    \centering
    \includegraphics[width=.24\textwidth]{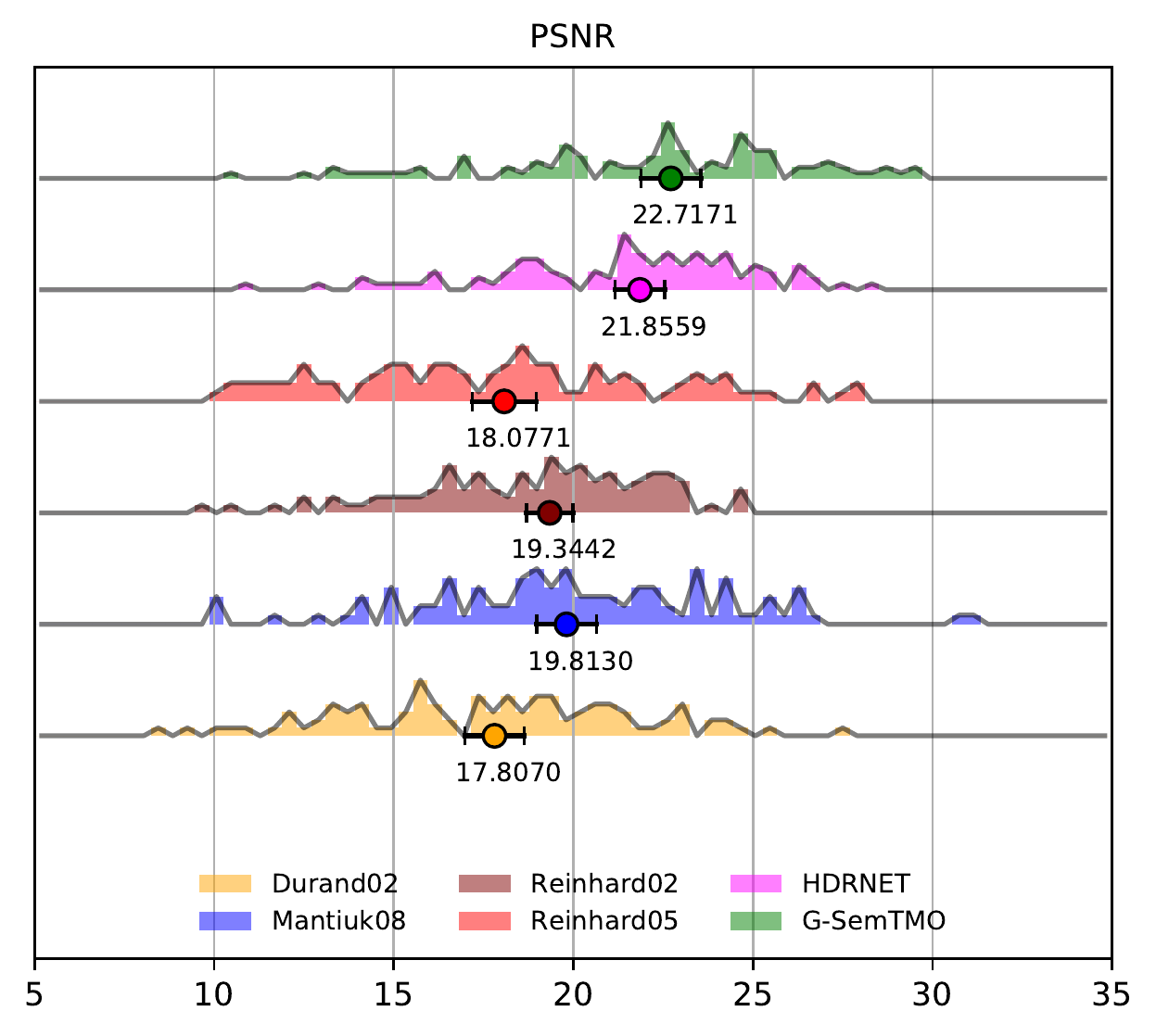} 
  \includegraphics[width=.24\textwidth]{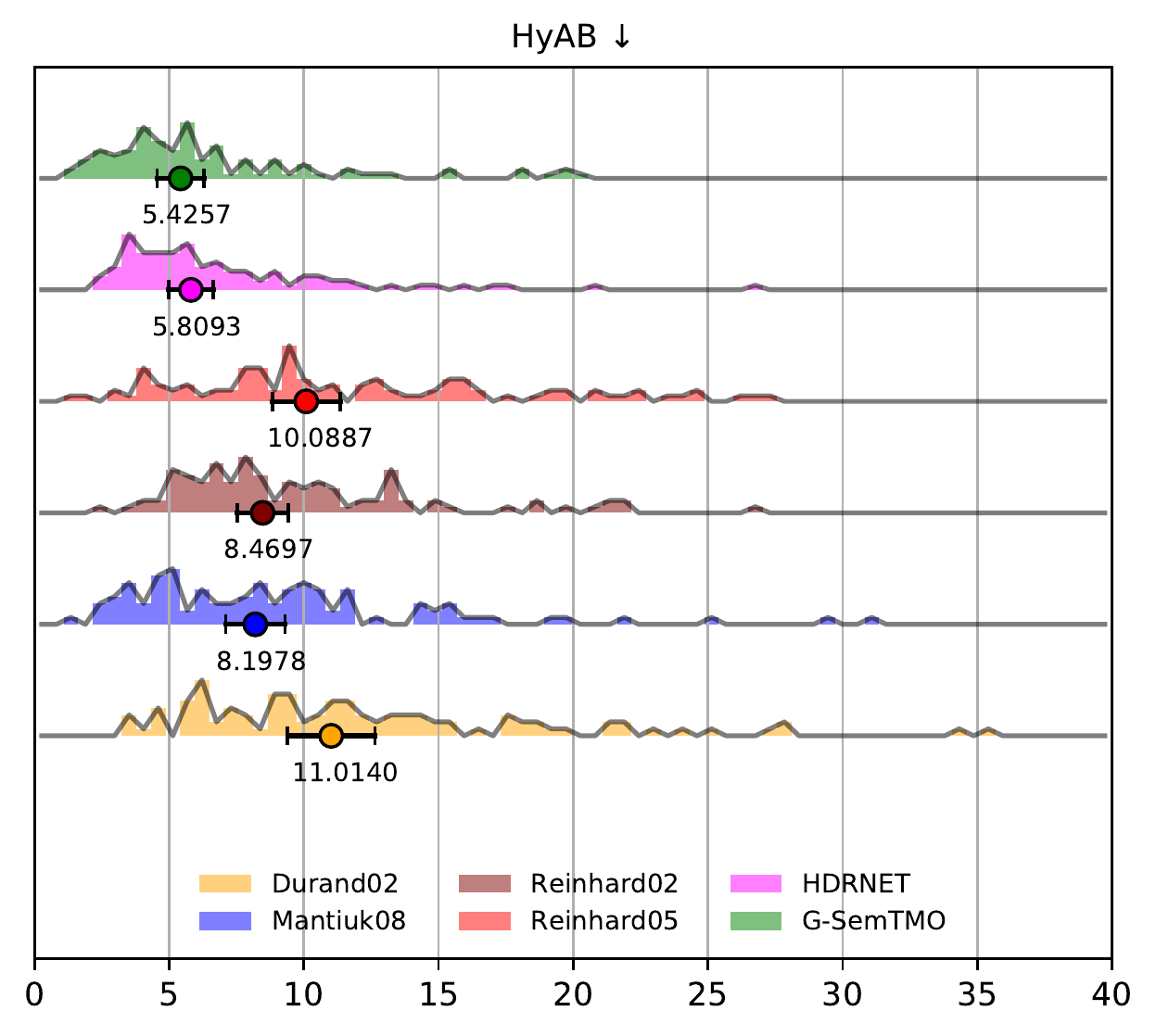} 
  \includegraphics[width=.25\textwidth]{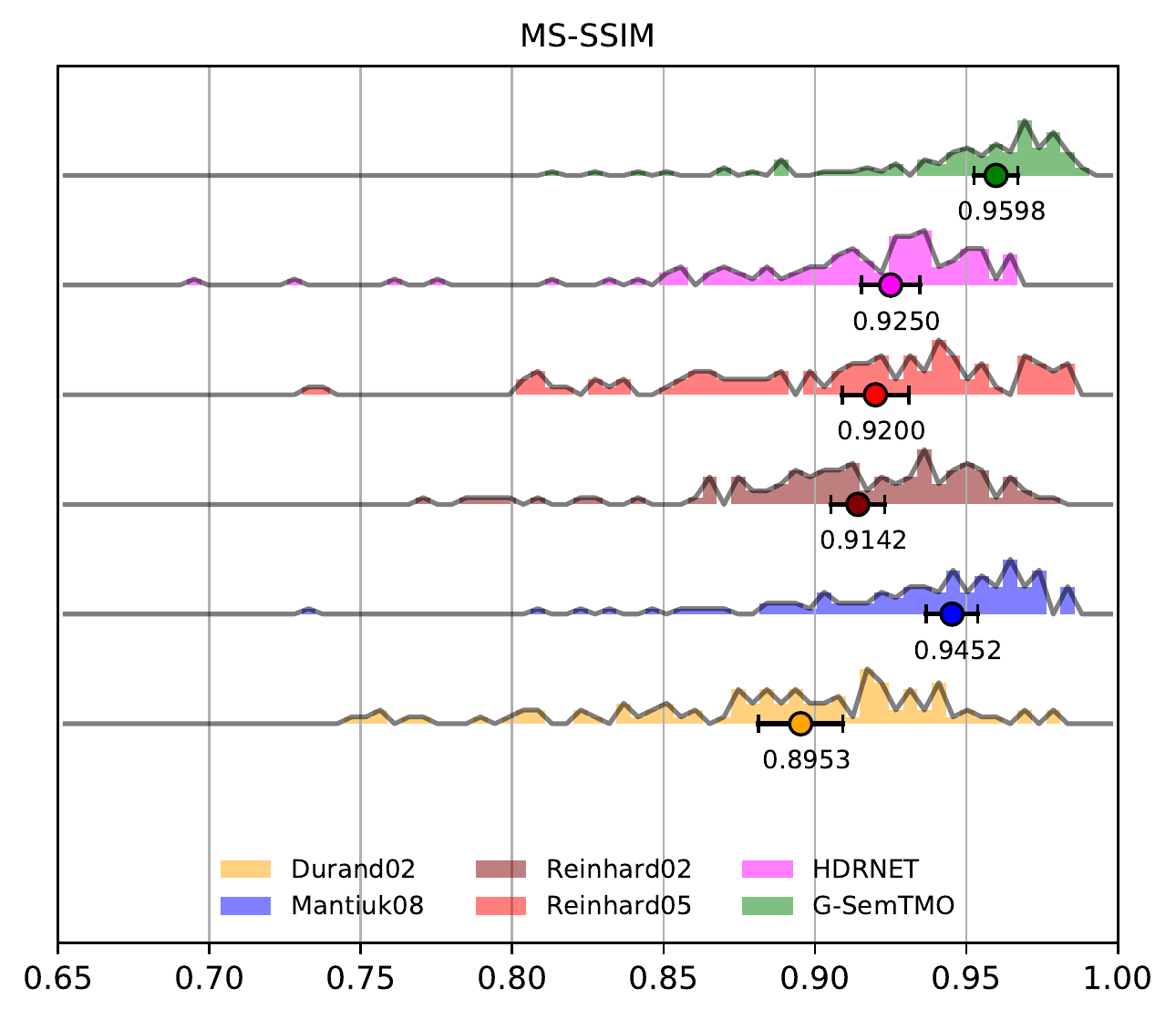} 
  \includegraphics[width=.24\textwidth]{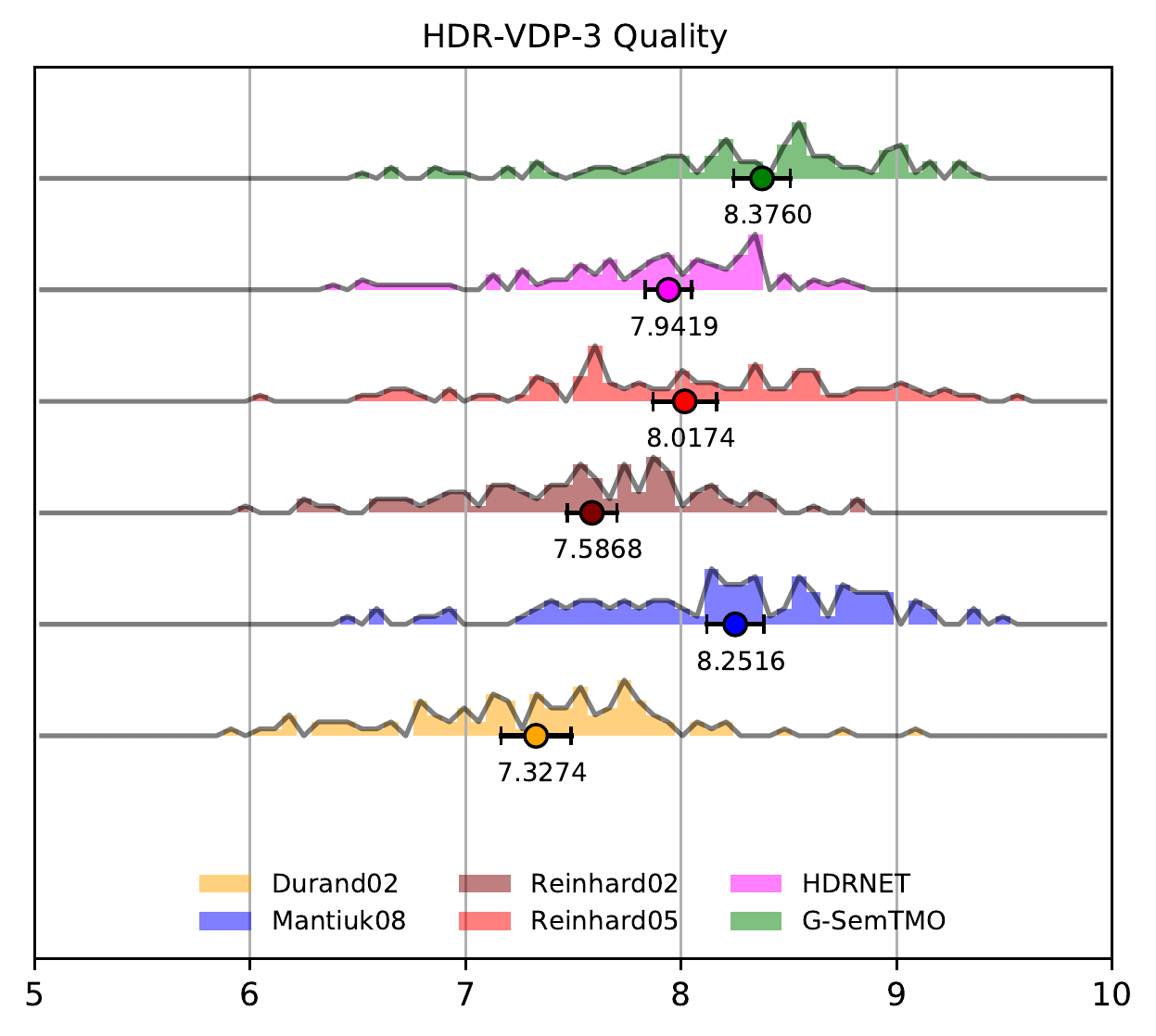} 
  \caption{\textit{Objective metrics score comparison}. 4 plots present 4 objective metrics (\textit{left to right}): PSNR, HyAB, MS-SSIM and HDR-VDP-3 Quality. Each plot presents 6 histograms of scores achieved by 6 TMOs: proposed G-SemTMO, HDRNET \cite{gharbi2017deep}, Photoreceptor TM \cite{reinhard2005dynamic}, Photographic TM \cite{reinhard2002photographic}, Display Adaptive TM \cite{mantiuk2008display} and Bilateral TM \cite{durand2002fast}. The median of each histogram is marked with a solid circle and a confidence interval of 95\%.}
  \label{fig:obj_metrics}
\end{figure*}

Another interesting observation can be made when analyzing the per-segment tone curves of G-SemTMO (the bottom row in \figref{comparison}). Each plot presents the tone curves predicted by G-SemTMO using the semantic hints per segment in a \textit{$\log_{10}$} space. We hypothesized that the neighborhood of semantic segments play a part in deciding the tonal adjustment inside the segment. Consequently, different neighborhood result in different tone curves for the same semantic label.

\figref{tc_explain} compares the graph representations of the semantic segments in two images A and B from \figref{comparison} (\textit{a4986} and \textit{a5000} respectively). Both images contain a large semantic segment annotated as \textit{vegetation} but the neighbors to \textit{vegetation} in A are different from B. Consequently, from \figref{comparison}, we observe that the tone curve for \textit{vegetation} is different in the two plots. Hence, we validate that the GCN learns the neighborhood information and predicts different hints for the same semantic label resulting in different tone curves. 

\figref{obj_metrics} shows the distribution of scores for aforementioned 4 objective metrics: PSNR, MS-SSIM, HDR-VDP-3 Quality and HyAB over 99 test images. For completeness and reference, the plot also includes the results for the 4 other traditional tone mapping operators apart from HDRNET and G-SemTMO, but as mentioned previously those operators were not trained to reproduce the results of expert~E. As reference, we use the version of the test images manually retouched by expert E from the FiveK dataset. Along with the histogram of observed metric scores, we plot the median metric scores for each TMO with an error bar denoting a confidence interval of $95\%$ of the median. The histograms confirm our subjective assessment of \figref{comparison}. We observe that across all objective metrics, the proposed G-SemTMO has a better median scores and produces results that are closer to the results of expert~E compared to the other TMOs. We notice that HDRNET results rival G-SemTMO closely in terms of colour difference (HyAB results) but there is a visible softness which is reflected in worse scores for more spatial metrics (MS-SSIM, HDR-VDP-3) sensitive to sharpness. Fortuitously, the display adaptive tone mapping also produces results that are close to the retouched images of expert~E. 
% The results are notably better in terms of MS-SSIM and HDR-VDP-3, which are sensitive to the lack of sharpness in the HDRNET results.

\begin{figure}[]
\centering
  \includegraphics[width=.82\linewidth]{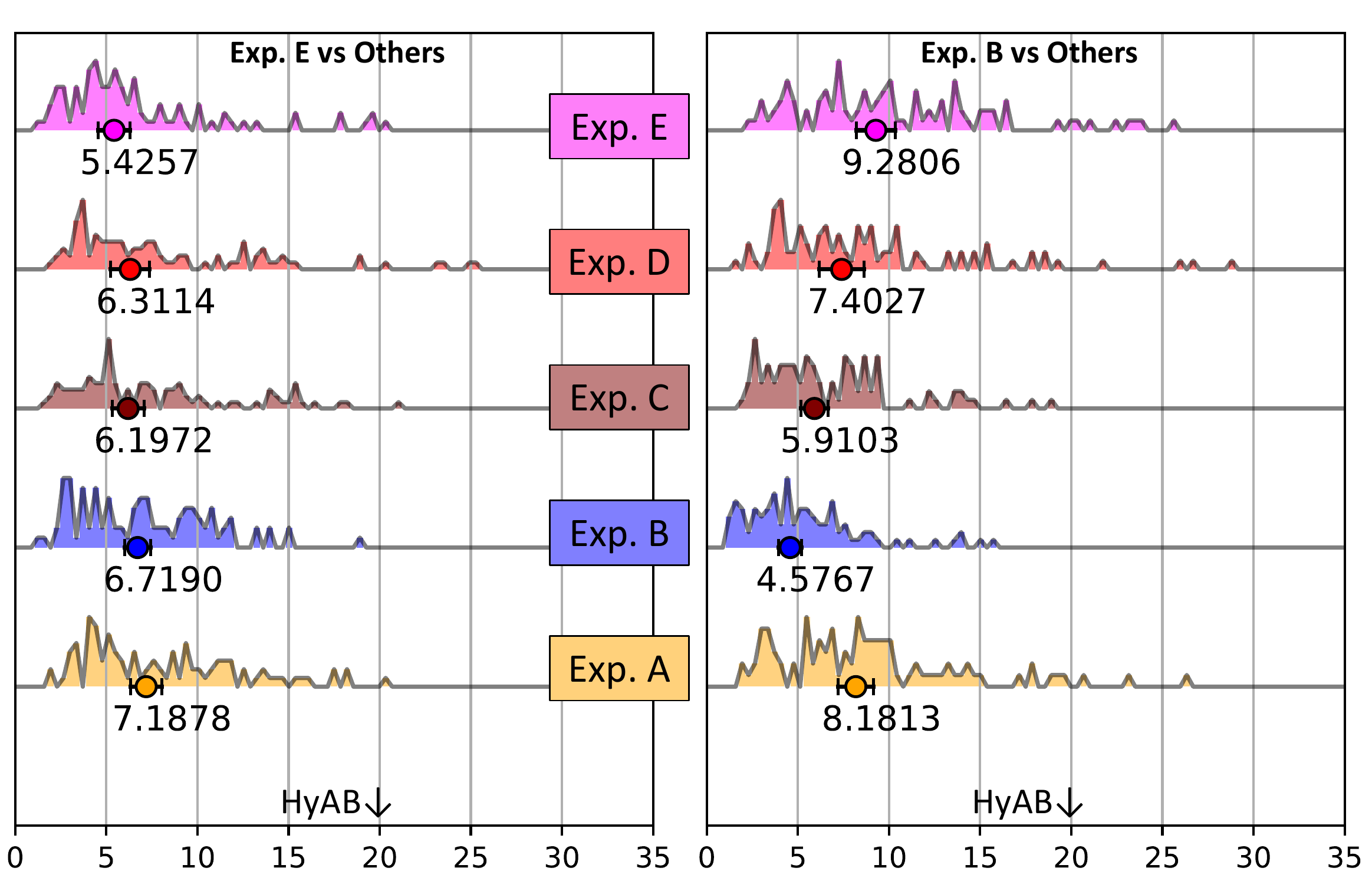}
  \caption{\textit{What you learn is what you get}: HyAB metric scores for expert E (left) and expert B (right) in comparison to the ground truth of other experts.}
  \label{fig:style differentiate}
\end{figure}

\begin{figure*}[]
\centering
  \includegraphics[width=.226\linewidth]{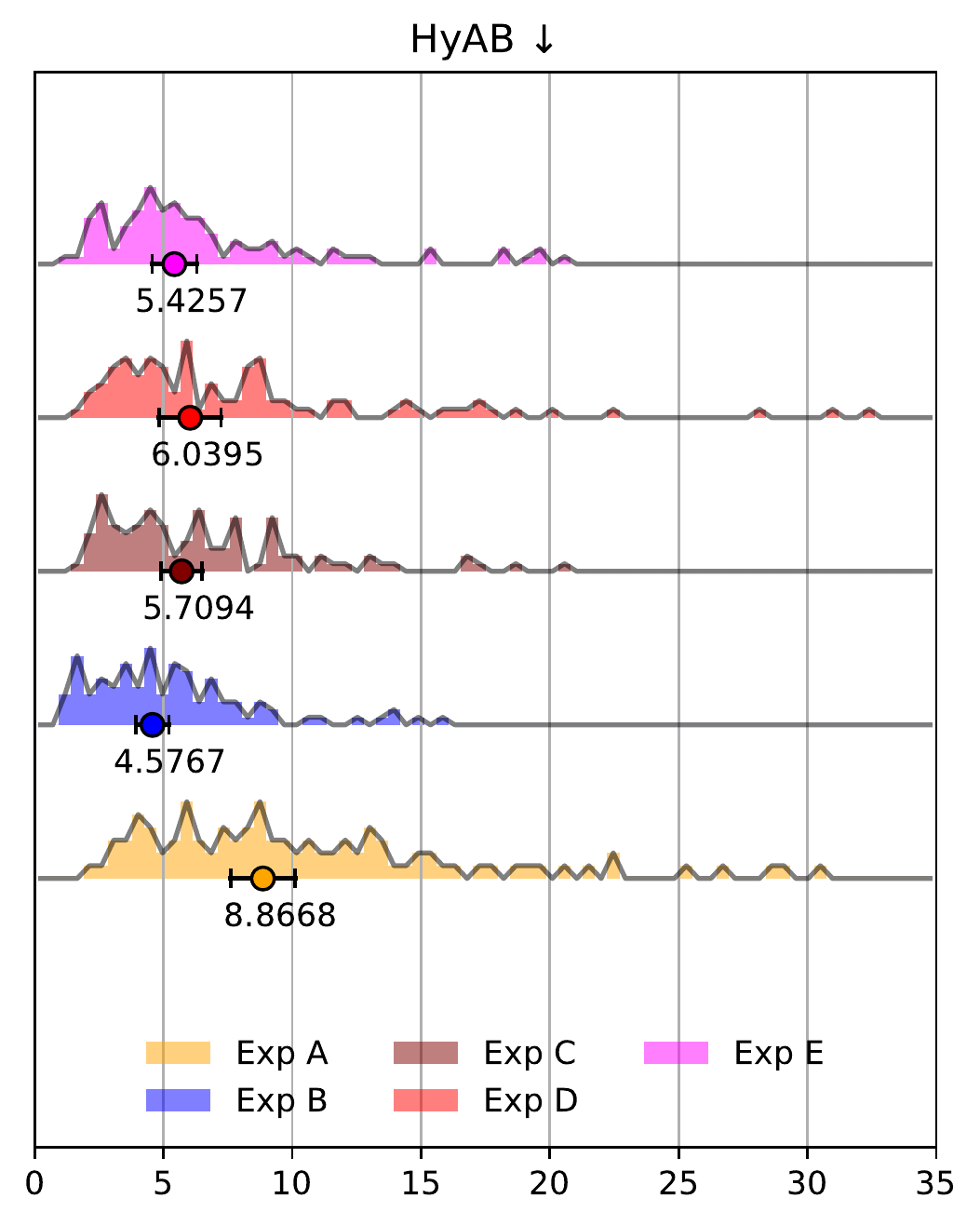}
  \includegraphics[width=.23\linewidth]{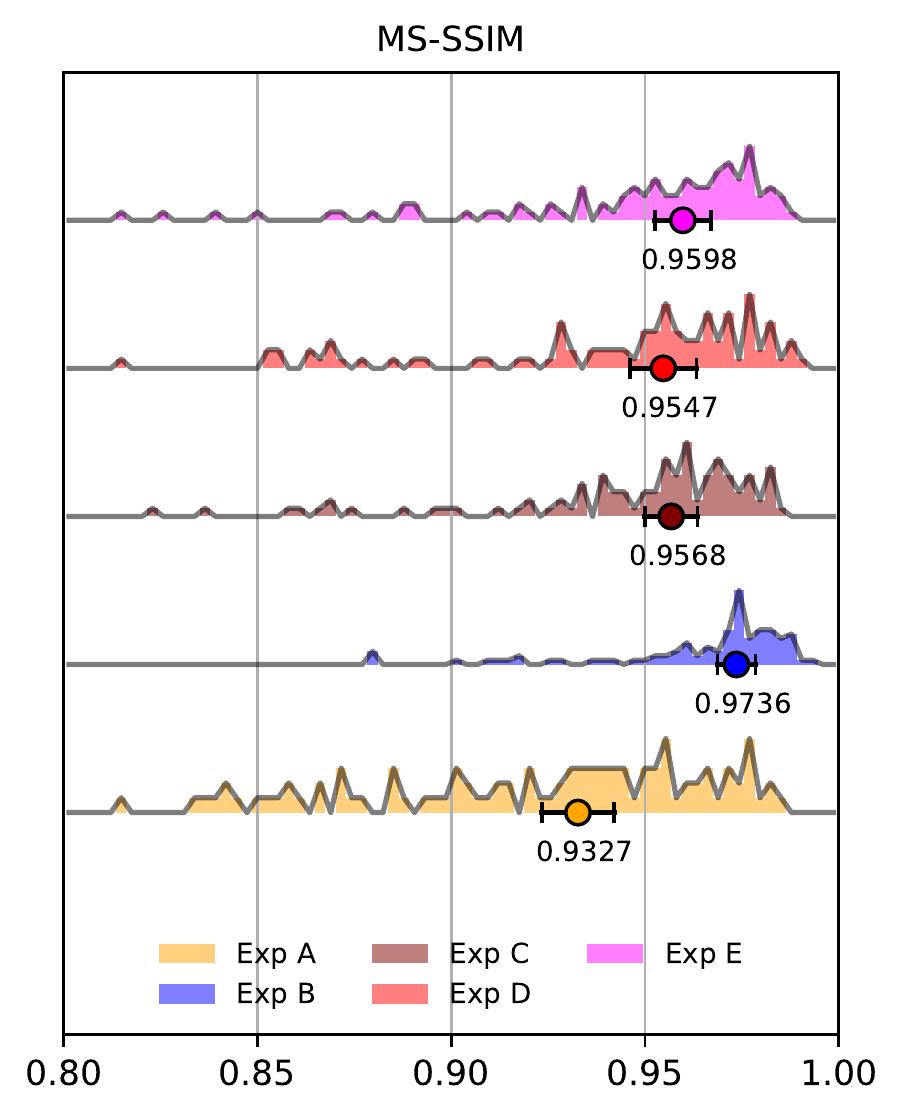}
  \includegraphics[width=.22\linewidth]{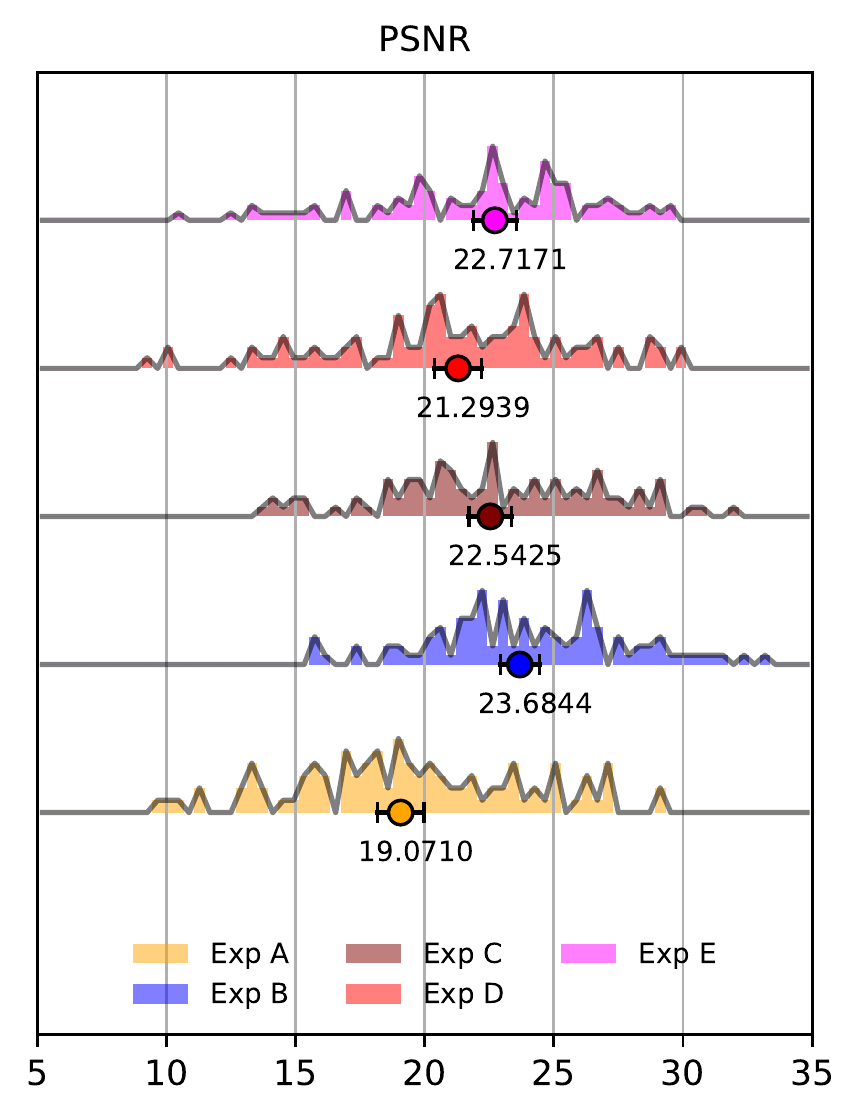}
  \caption{Metric score distributions for HyAB, MS-SSIM and PSNR for networks trained over 5 experts individually. The plots show the histograms of scores along with medians and its 95\% confidence intervals.}
  \label{fig:5-experts}
%   \vspace{2mm}
%   \includegraphics[width=.24\textwidth]{Illustrations/metrics/PSNR_median.pdf} 
%   \includegraphics[width=.24\textwidth]{Illustrations/metrics/HyAB_median.pdf} 
%   \includegraphics[width=.25\textwidth]{Illustrations/metrics/Msssim_median.pdf} 
%   \includegraphics[width=.24\textwidth]{Illustrations/metrics/hdrvdpq_median.pdf} 
%   \caption{\textit{Objective metrics score comparison}. 4 plots present 4 objective metrics (\textit{left to right}): PSNR, HyAB, MS-SSIM and HDR-VDP-3 Quality. Each plot presents 6 histograms of scores achieved by 6 TMOs: proposed G-SemTMO, HDRNET \cite{gharbi2017deep}, Photoreceptor TM \cite{reinhard2005dynamic}, Photographic TM \cite{reinhard2002photographic}, Display Adaptive TM \cite{mantiuk2008display} and Bilateral TM \cite{durand2002fast}. The median of each histogram is marked with a solid circle and a confidence interval of 95\%.}
%   \label{fig:obj_metrics}
  \vspace{2mm}
  \includegraphics[width=.84\linewidth]{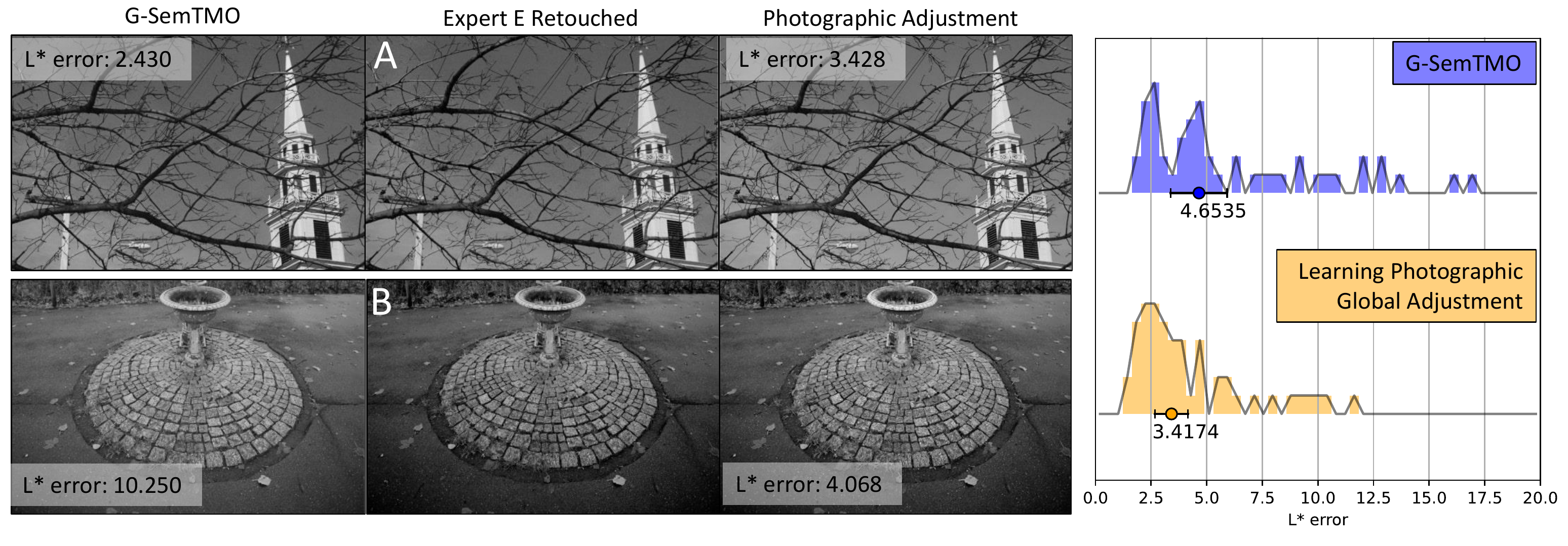}
  \caption{Comparing L* prediction: G-SemTMO and network trained by Bychkovsky et al.~on Adobe FiveK dataset to learn photographic global adjustment \cite{fivek}. Bychkovsky et al.~only predict luminance adjustment and copy the colour profile produced by the expert. G-SemTMO, trains for a much more complex end-to-end mapping of both colour and luminance. The plot shows L* error (from CIELAB space) histograms for 50 test images predicted by either method. Additional images provided with supplementary material.}
  \label{fig:compare_fivek_adjus}
  \vspace{2mm}
  \includegraphics[width=.84\linewidth]{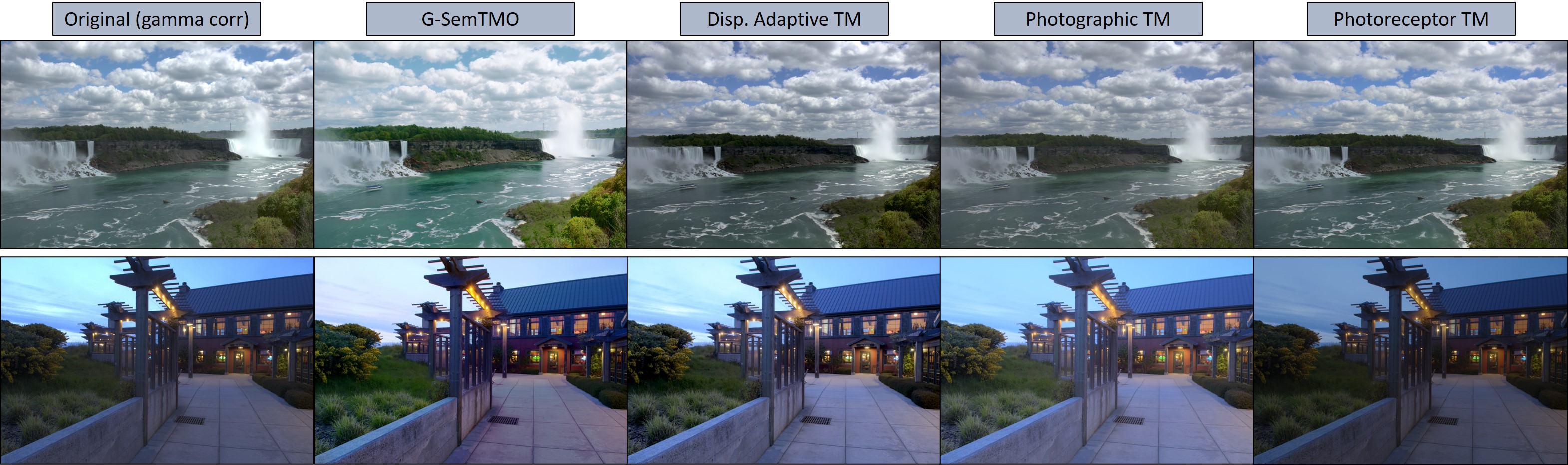}
  \caption{Using G-SemTMO (trained on expert E) outside the box. Images from Fairchild HDR \cite{Fairchild} dataset tone mapped with Disp. Adapt. TM, Photographic TM and Photoreceptor TM and proposed G-SemTMO. Additional images provided with supplementary material. }
  \label{fig:fairc}
\end{figure*}

\subsection{Training for other expert photographers}
\label{sec:other_experts}
We trained our network over the same set of training images for the 4 other expert photographers in the FiveK dataset and validated the results over the 99 test images. We use the same hyper-parameters for training as in \secref{training_details}. We observe that there are inconsistencies among the tonal adjustments provided by the experts in the FiveK dataset as a result of which learning tone mapping becomes harder.

Consequently, to validate that our network can differentiate between the styles of each expert and learn tonal adjustment specific to the expert, we compare the prediction of G-SemTMO trained for a particular expert to the other expert ground truth. \figref{style differentiate} shows the performance of networks trained over expert E and B with HyAB metric. We observe that results predicted by network trained over E is closest to the ground truth E than others for 99 images. The same holds true for network trained over expert B. This concludes that the parameters learnt by the network are specific to the expert trained.

Gharbi et al.~\cite{gharbi2017deep} mention that HDRNET could learn the adjustments made by expert B better. We also notice that our training could learn and infer better for expert B, as validated by the objective metric scores in \figref{5-experts}. Subjectively analysing the enhancements, we find the adjustments made by expert A to be the most inconsistent.

\subsection{Comparing to Photographic Global Tonal Adjustment}
Bychkovsky et al.~\cite{fivek} used the FiveK dataset to train over expert E to learn specific adjustment style for images. However, they predict the luminance adjustment for test images only and apply the colour profile as curated by expert E for each test image. To maintain fairness we compare the luminance predicted  by G-SemTMO to those of Bychkovsky et al.~over 50 common images among their and our test set. We observe in \figref{compare_fivek_adjus} that G-SemTMO predicts luminance closer to expert for image A but performs worse for image B. Overall, Bychkovsky et al.~ have a statistical advantage for better luminance prediction as we observe from the L* error histogram in the same figure. However, it should be noted that G-SemTMO is trained on the end-to-end problem and learns a far more complex mapping than just luminance adjustment.

\subsection{G-SemTMO outside the box}
We can consider an expert photographer's process of retouching to be a manual analogy to tone mapping. Since we train G-SemTMO to mimic an expert's style, we test how G-SemTMO generalizes to other dataset and whether it can recreate the retouching style of expert~E. We conduct an aesthetic evaluation of G-SemTMO as a standalone tone mapping operator. We choose Fairchild's HDR dataset \cite{Fairchild} and compare the results of G-SemTMO against 3 classical TMOs. \figref{fairc} presents tone mapped results of 2 images and shows that G-SemTMO can produce aesthetic results as a standalone operator. We exclude HDRNET tone mapping results from this comparison because of visible color artifacts in its prediction.

\section{Learning Local Image Enhancements}
It can be argued that the tonal adjustments created by the expert photographers for the FiveK~\cite{fivek} dataset is global in nature. The photographers had access to limited tools and sliders from the Adobe Lightroom photo-retouching application. Although, the sliders can effect non-linear adjustments, they are not as local as using brushes and radial/gradual filters to modify images. It is important to validate the performance of G-SemTMO in learning local tonal adjustments. Consequently, we present a locally enhanced dataset of HDR images, \textit{LocHDR}. We train over our dataset and conduct ablation studies to confirm whether graph convolution manages to learn tone modifications closer to the reference.
 
\subsection{Local HDR Dataset - LocHDR}
We filter the images from FiveK based on their dynamic range. We compute dynamic for images from FiveK range as the logarithm of the ratio of the $99^{th}$ and $1^{st}$ percentile of observed luminance and empirically put a threshold of $2.2$ to filter out images. Furthermore, to emphasize on local changes, we filter images based on the number of semantic segments and choose images with at least 3 unique semantic labels. Based on our criteria we compile a subset of $781$ HDR images. We hire an expert photo-retoucher (henceforth referred to as Expert I) who is tasked to apply corrections to the LocHDR dataset using Adobe Lightroom application with emphasis on using brushes and spatial filters. Only the sliders in the Tone section --- \textit{Exposure, Contrast, Highlights, Shadows, Whites} and \textit{Blacks} are used and no auto-enhancement or colour, noise, detail adjustments are made.

\begin{figure}[t]
\centering
  \includegraphics[width=.94\linewidth]{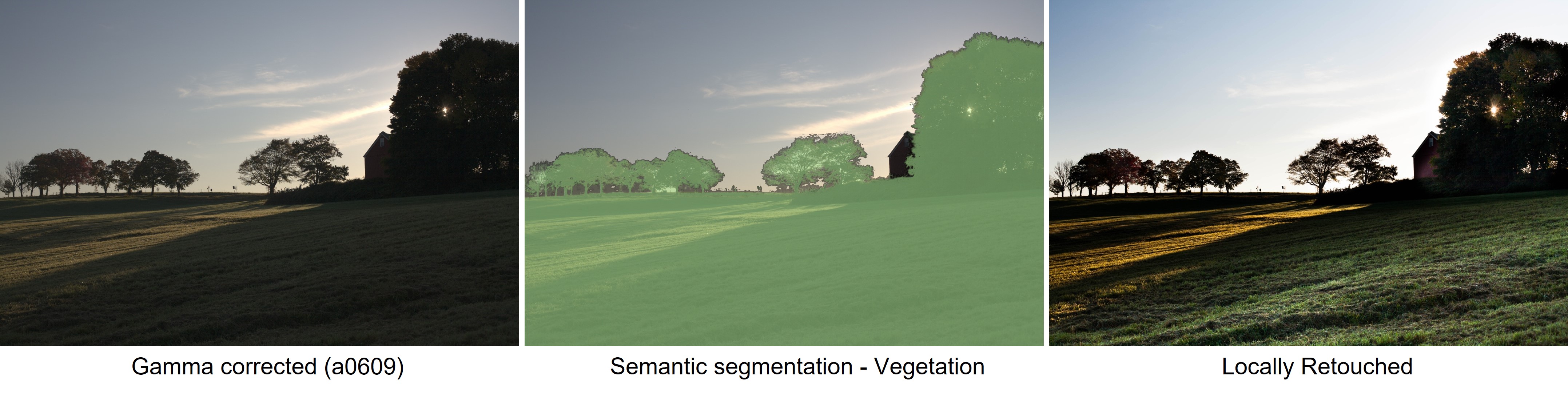}\\
  \includegraphics[width=.94\linewidth]{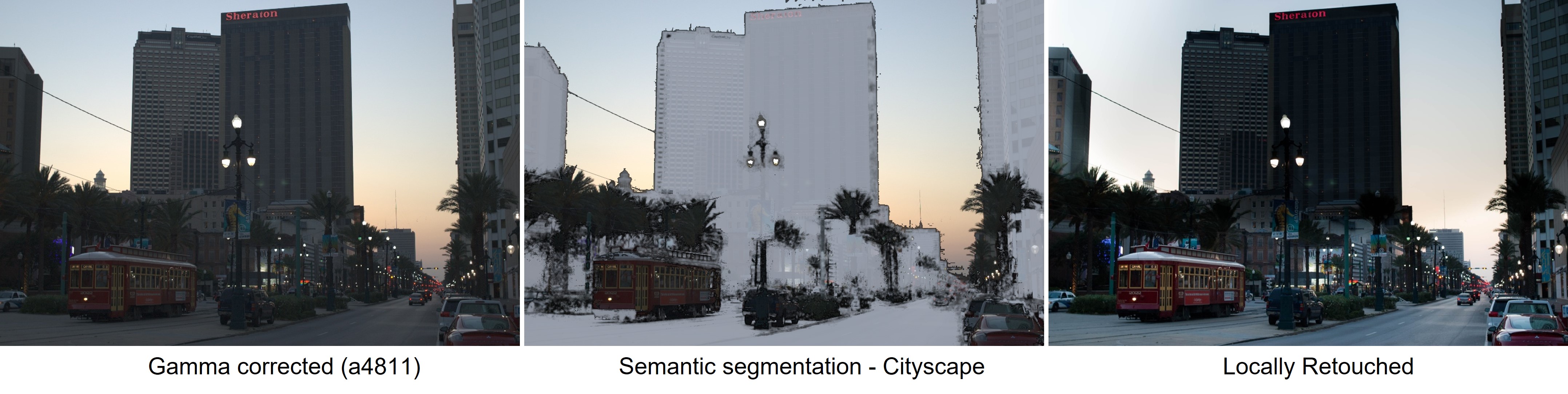}\\
  \includegraphics[width=.94\linewidth]{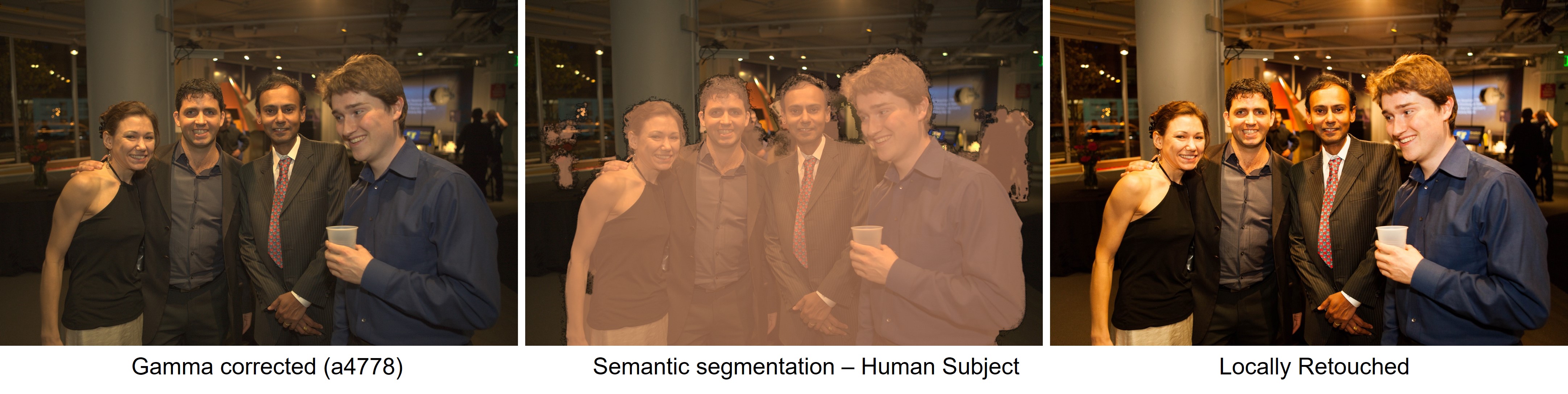}\\
  \includegraphics[width=.94\linewidth]{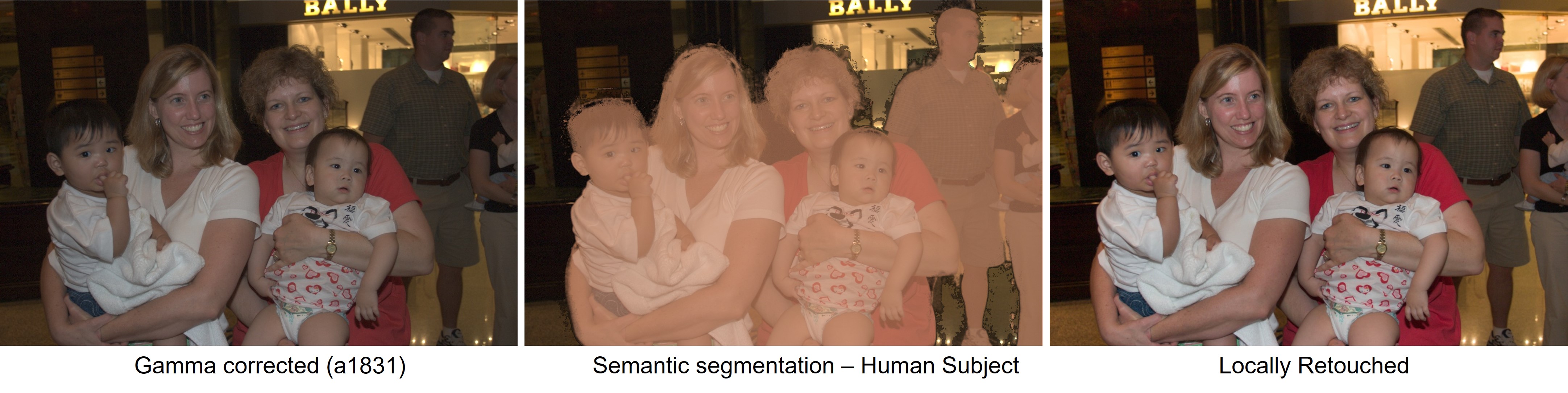}
  \caption{Inconsistencies in local enhancements. From Left-Right: the gamma corrected image, chosen semantic class and locally enhanced image. We can observe noticeable variation of local contrast inside masked semantic classes \textit{Vegetation} and \textit{Cityscape} for images a609 and a4811. Furthermore, we notice contrast variations across images a4778 and a1831 for the same semantic class \textit{Human Subject}. Image a4778 has a high local and global contrast making it the enhancement perceptually different that image a1831.}
  \label{fig:inconcistency}
\end{figure}

On closer observation and investigation we uncover that manual enhancement using local tools leads to tonal enhancement inconsistencies in the LocHDR Dataset. \figref{inconcistency} illustrates how local enhancements across and within images for the same semantic label vary. On image `a0609' we observe that parts of the \textit{Vegetation} segment has received inconsistent exposure gains. Similarly, parts of the \textit{Cityscape} segment in image `a4811' are well exposed whereas parts of the building are darkened. Comparing the enhanced image `a4778' to `a1831' we observe that the former appears 'punchy' with heightened contrast on the \textit{Human Subject}.

Learning tone mapping from reference pairs through a data-driven approach is similar to learning individual retouching styles. Our expert photo-retoucher does not have access to the semantic masks which the G-SemTMO uses to learn local enhancements. Hence, it becomes challenging for the network to train over an image dataset with local enhancement inconsistencies. The network fails to generalize and converge on the style of adjustment. Previous work on FiveK dataset~\cite{fivek} mentions the inconsistencies in retouches~\cite{gharbi2017deep, yan2016automatic} and the use of data splits such as `Random 250' and `High Variance 50'~\cite{gharbi2017deep, hwang2012context, yan2016automatic} for analysis. We also decide to split our LocHDR based on consistency in style. 

\subsection{Style-specific High Contrast Enhancement - HC200}
In order to evaluate whether the presence of semantic graphs helps G-SemTMO learn local enhancements better, we present a final dataset of \textit{High Contrast 200} images. We filter the LocHDR dataset based on the perceivable contrast effected by Expert I to maintain consistent enhancement in training set.

Measuring the perceptual contrast or how `contrast-y' or `punch-y' an image appears is a challenging task. Inspired by multi-level approaches in entropy computation~\cite{mle} and structural similarity measures~\cite{wang2003multiscale}, we present our own approximation of a multi-level contrast measure. Multi-level contrast follows a multi-grid approach where at each level $n$, the full resolution image is divided into a grid of $n\times n$ patches and patch-specific variance of pixel intensity is computed. The contrast for level $n$ is the square root of the mean variance. The final multi-level contrast measure is computed as the mean of level-specific contrast scores thereby capturing the global as well as local contrast variations:
\begin{equation}
\displaystyle
    C_\text{ML} = \frac{1}{n}\sum^{n}_{i=1}\left( \sqrt{\frac{\sum^{i^2}_{p=1}Var_p}{{i^2}}} \right),
\end{equation}
where $n$ is the number of levels, and $p$ is the index of patches in a level from 1 to $n\times n$. We empirically set $n=5$ for our contrast estimation. We choose the 200 images with highest contrast measure. On visible subjective assessment, we can confirm that the HC200 subset mostly contains the high contrast `punchy' images from LocHDR.

\begin{figure*}[b]
    \centering
    \footnotesize
    \setlength\tabcolsep{1.5pt}
    \begin{tabular*}{\textwidth}{ccccc}
    % \toprule
    Gamma Corrected & 3D LUT-Local - LocHDR & G-SemTMO - LocHDR & G-SemTMO - HC200 & Ground Truth\vspace{2mm} \\ 
   
      \includegraphics[width=.19\linewidth]{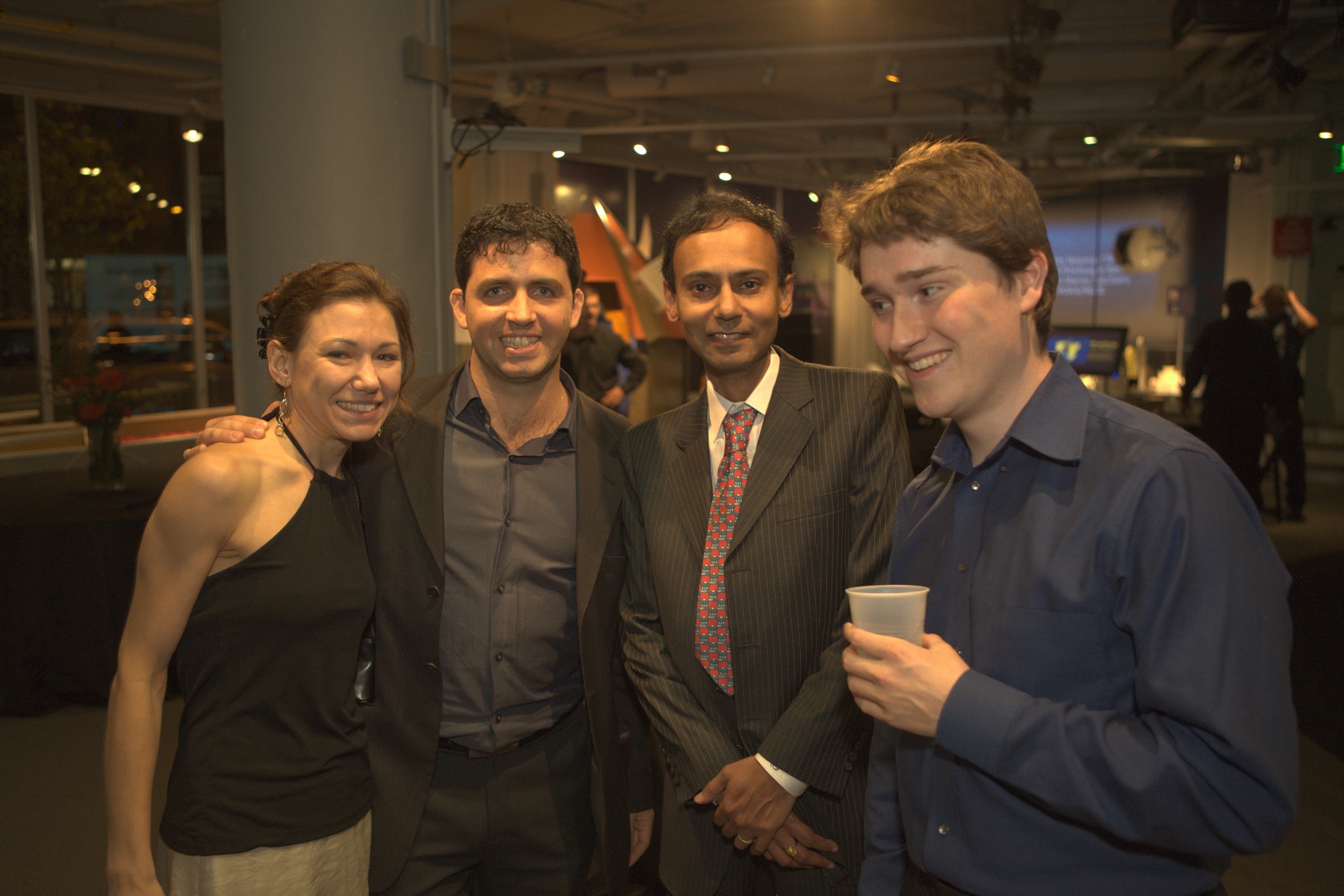} & \includegraphics[width=.19\linewidth]{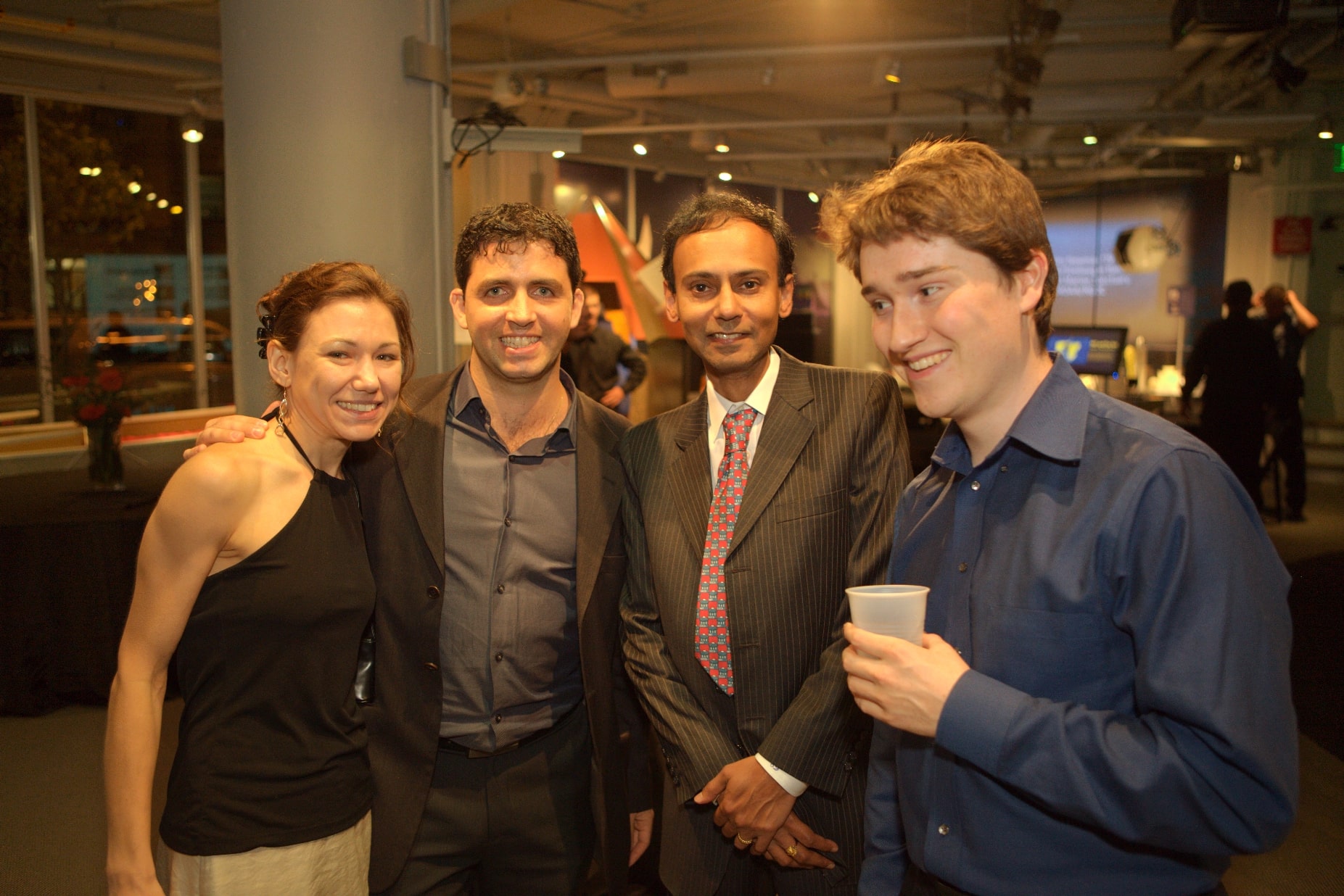} &  \includegraphics[width=.19\linewidth]{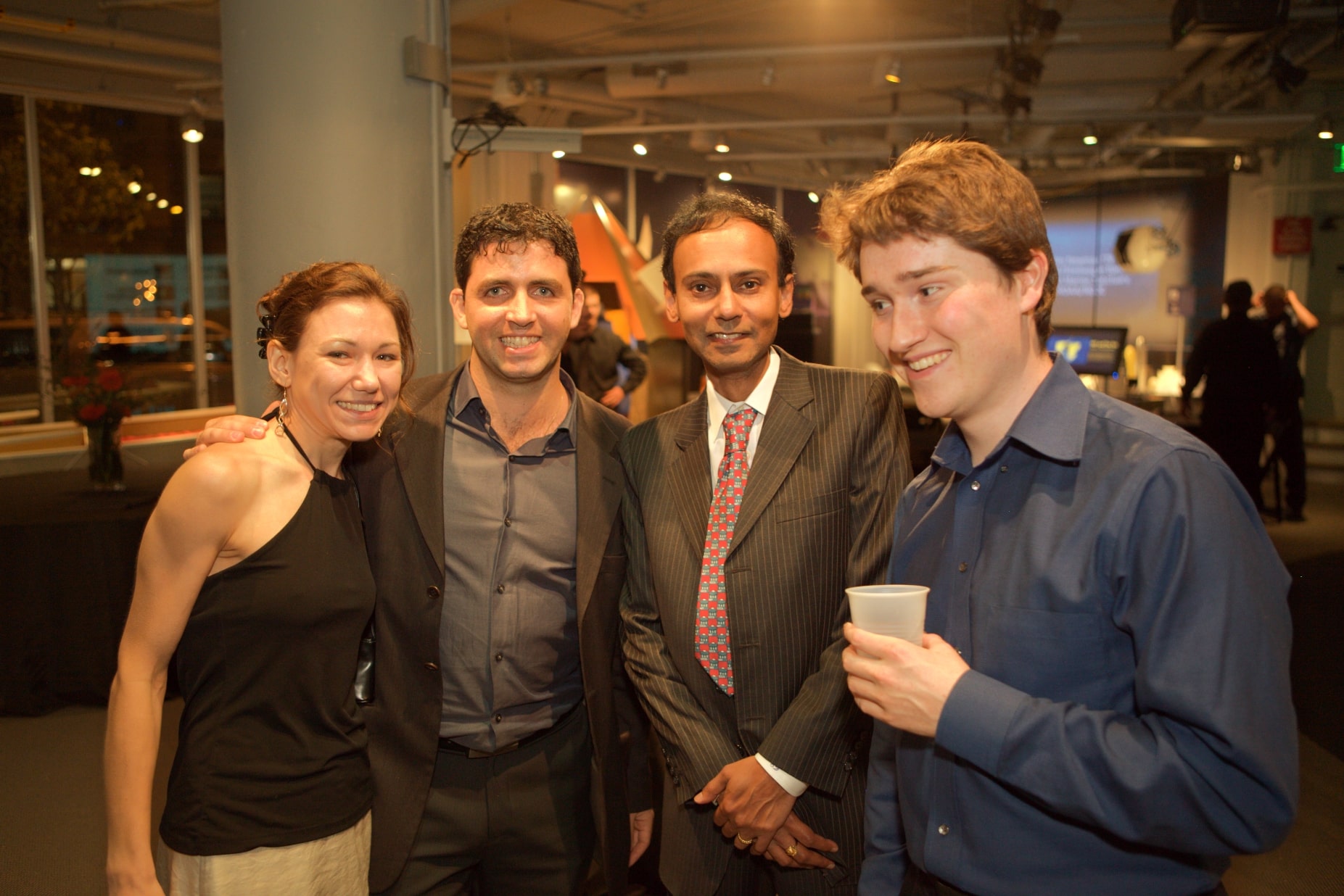} & \includegraphics[width=.19\linewidth]{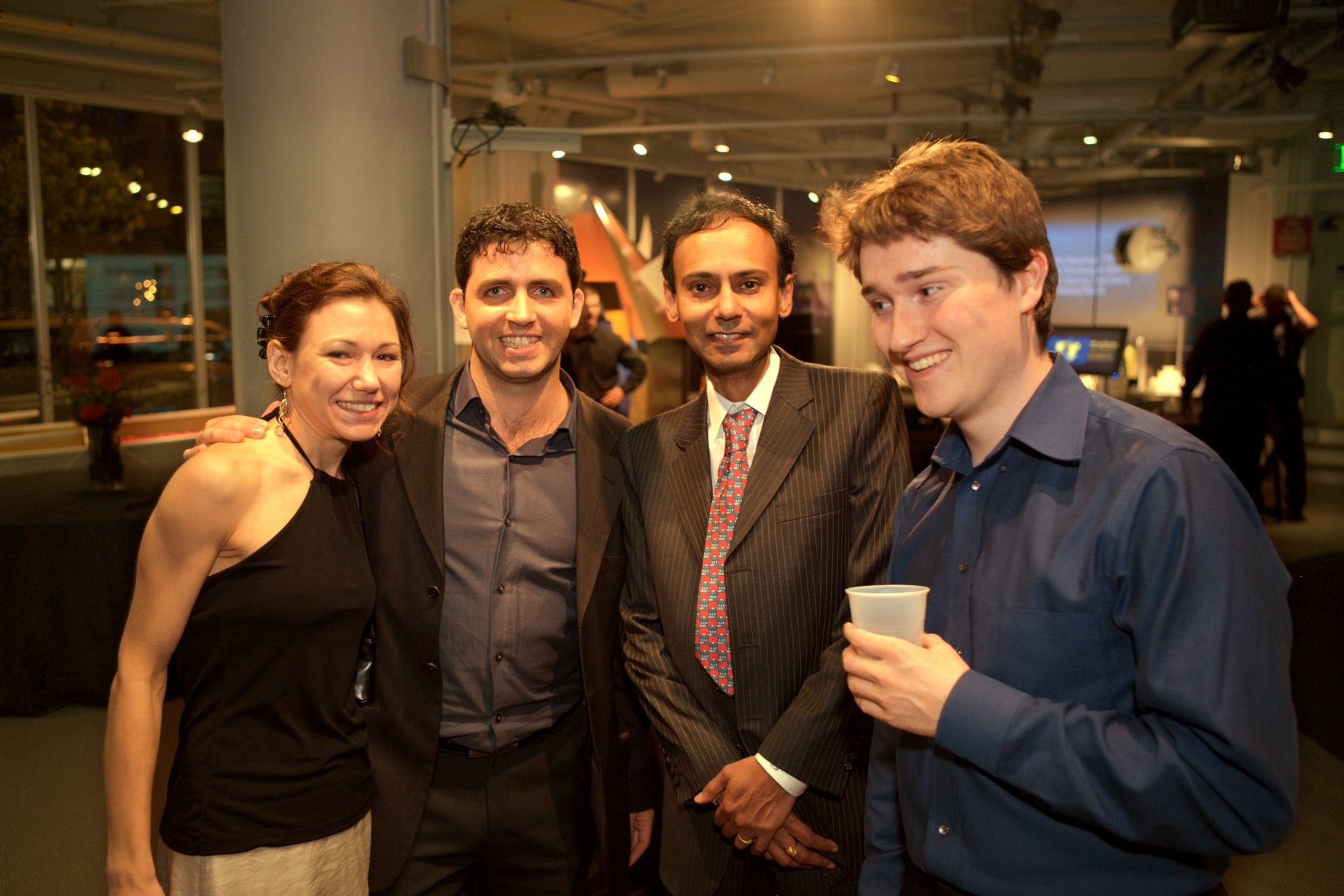} &  \includegraphics[width=.19\linewidth]{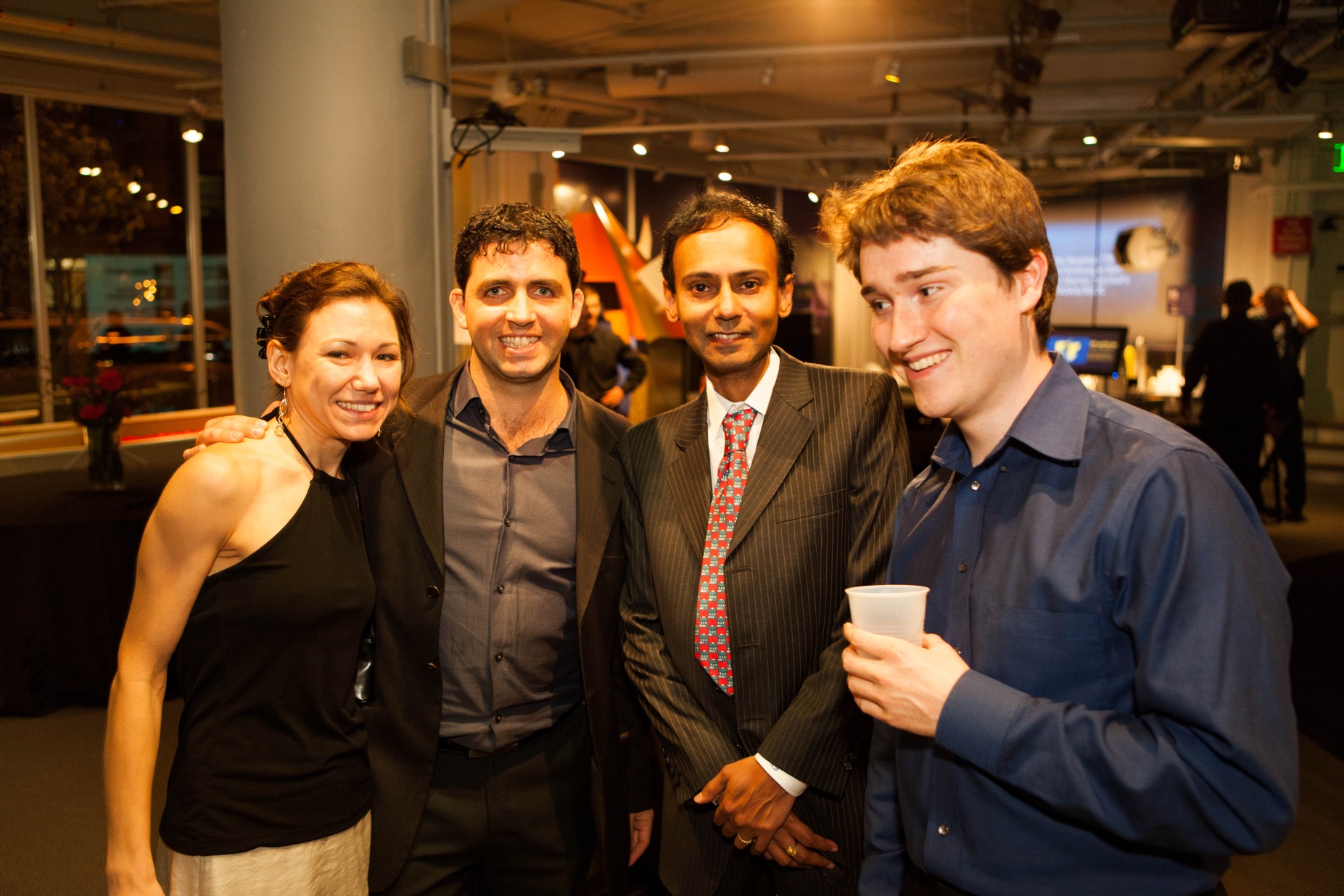} \\ 
      a4778 & HyAB: 4.30 \; PSNR: 24.58 & HyAB: 4.66 \; PSNR: 24.98 & \textbf{HyAB: 3.13 \; PSNR: 26.14}&  \vspace{1mm}\\
      \includegraphics[width=.19\linewidth]{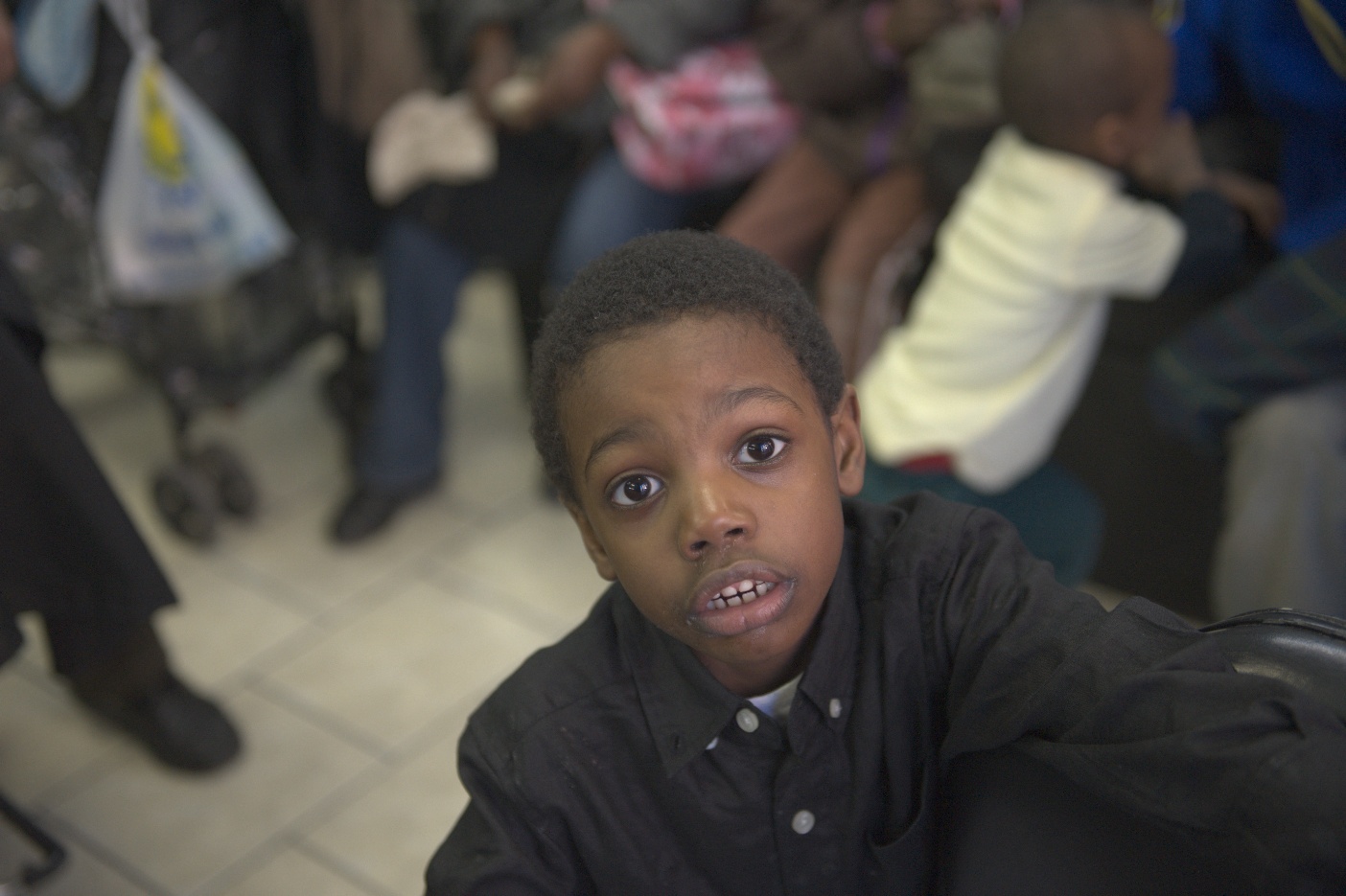} & \includegraphics[width=.19\linewidth]{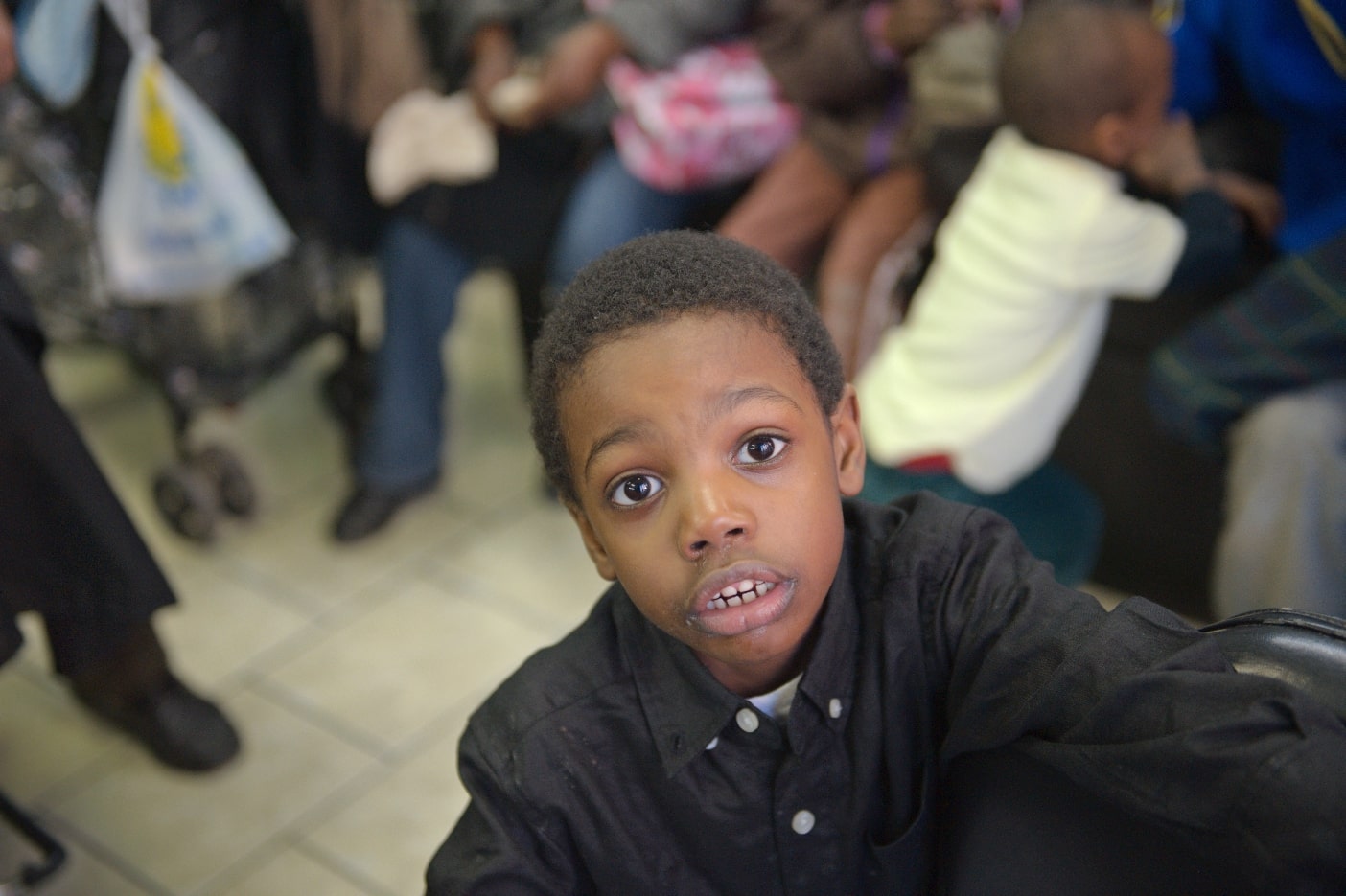} & \includegraphics[width=.19\linewidth]{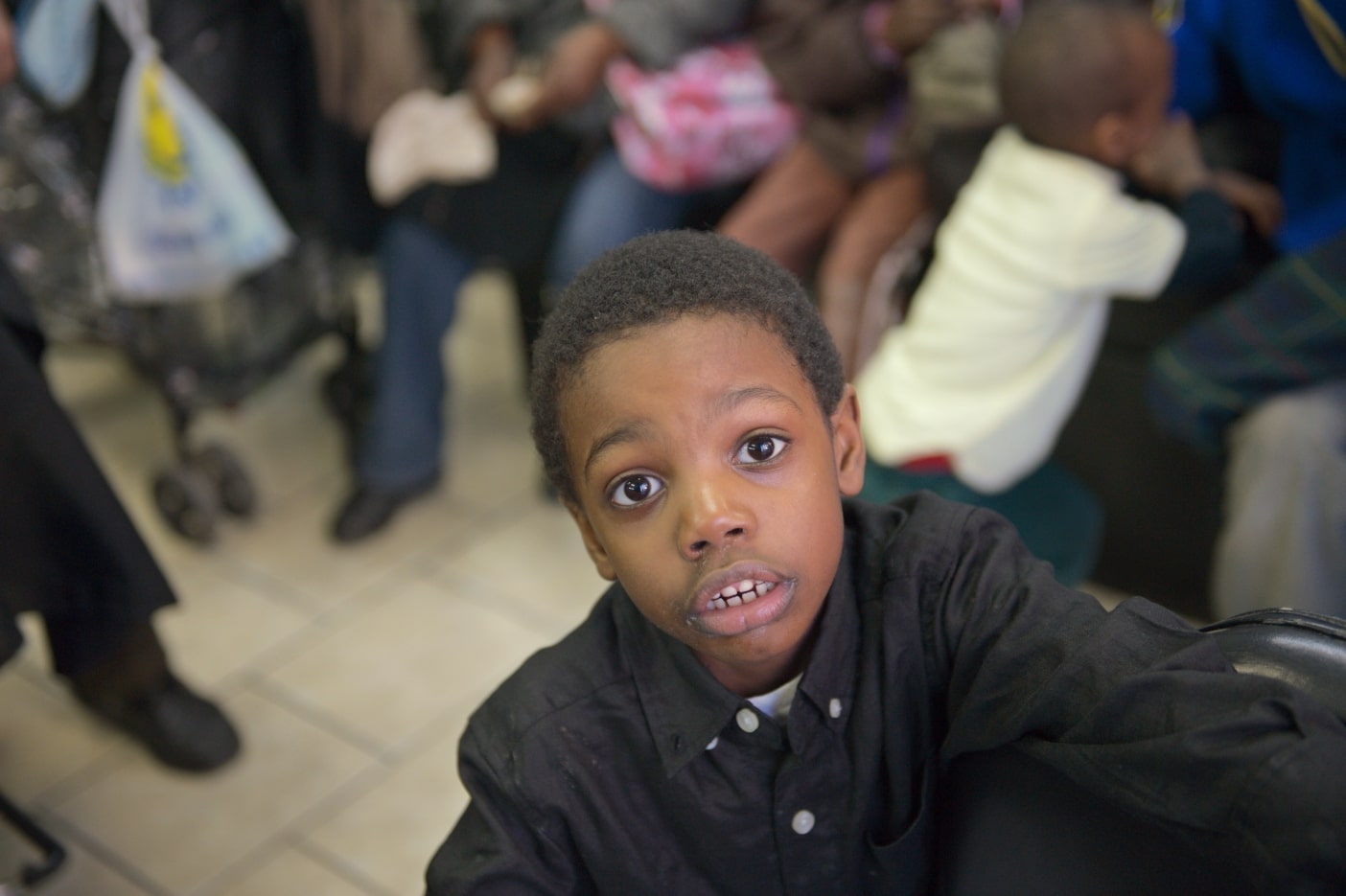} &
      \includegraphics[width=.19\linewidth]{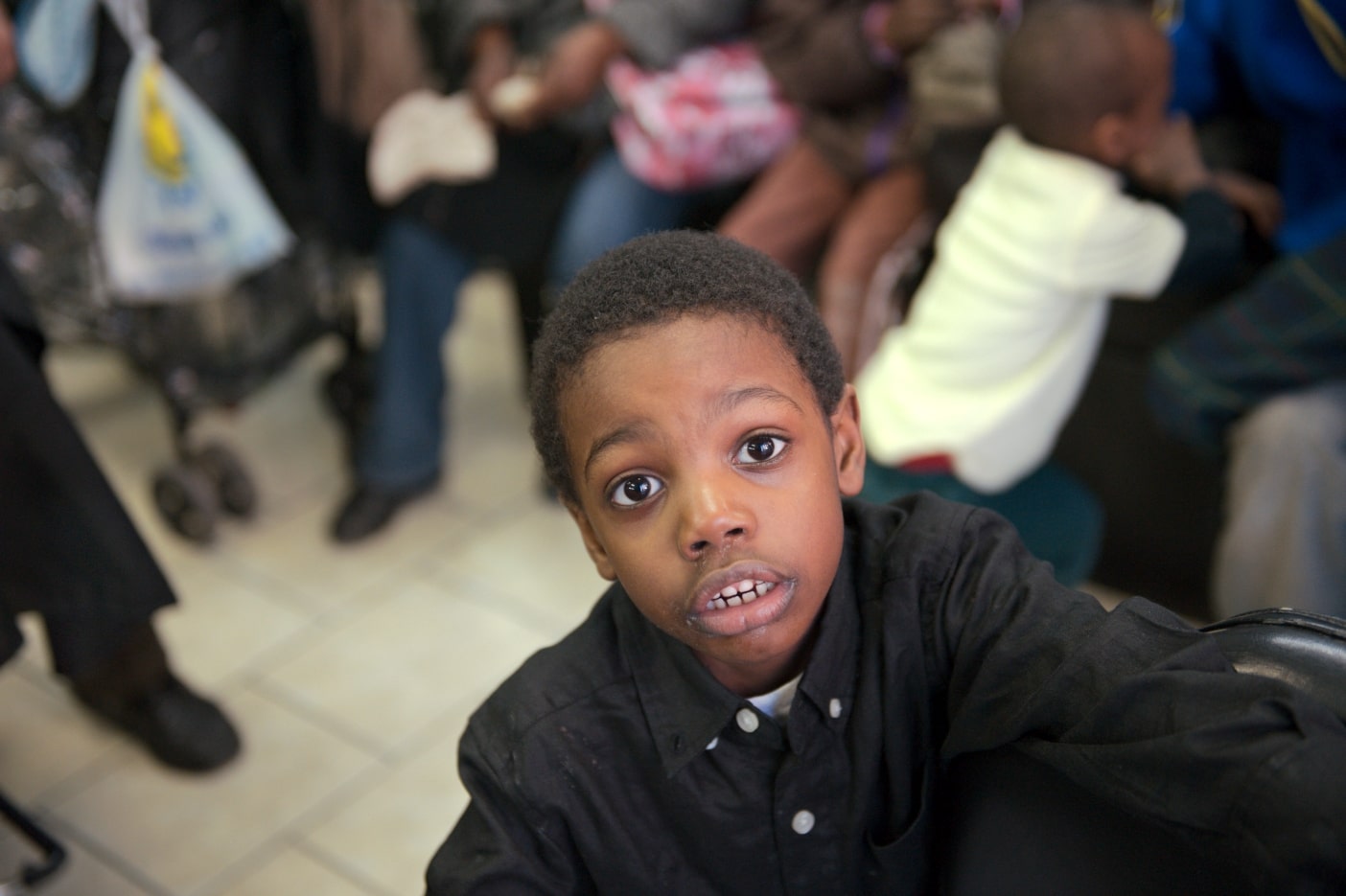} & \includegraphics[width=.19\linewidth]{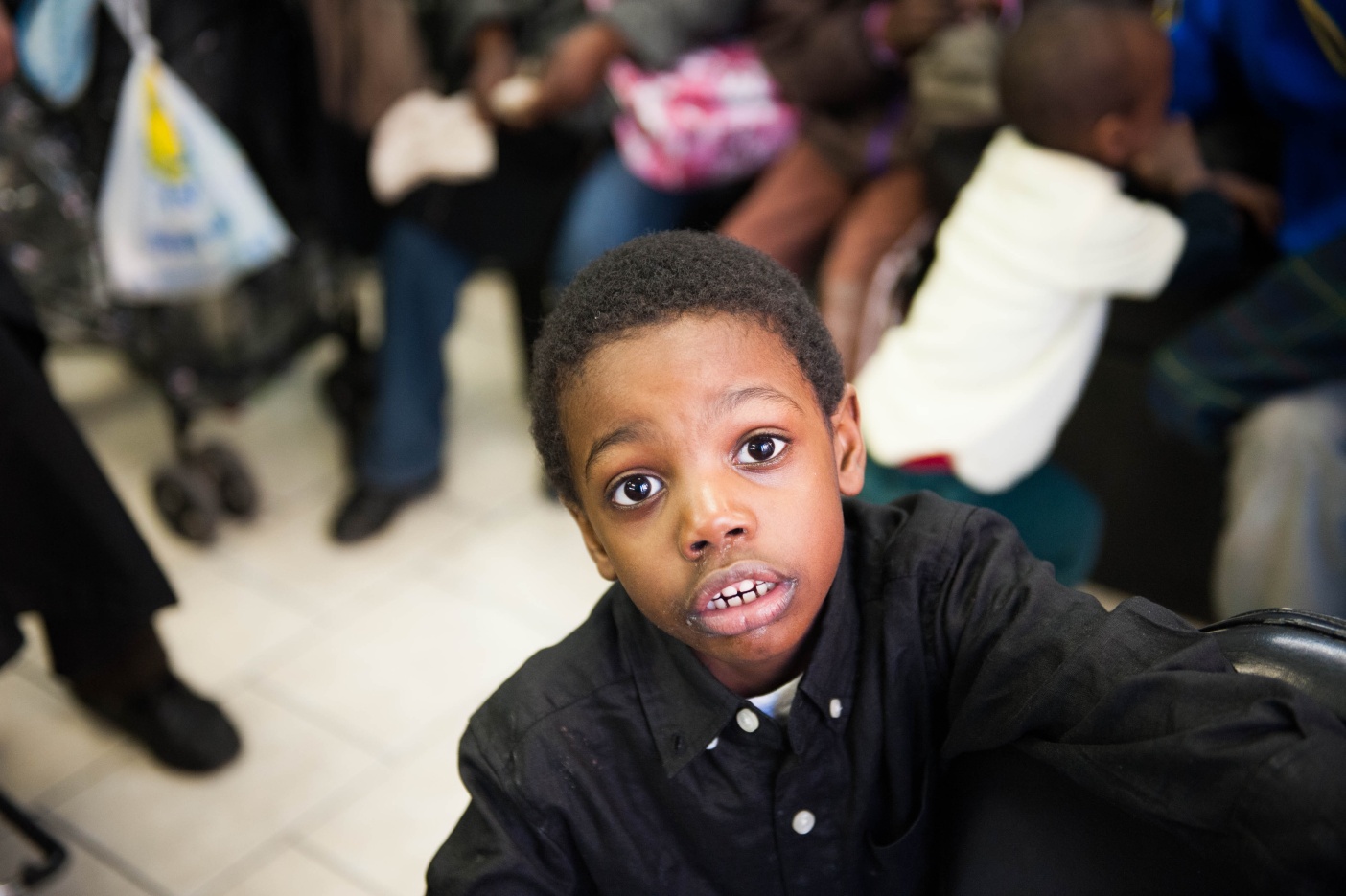} \\ 
      a4866 & HyAB: 8.02 \; PSNR: 19.58 & HyAB: 7.81 \; PSNR: 19.72 & \textbf{HyAB: 4.98 \; PSNR: 22.86} & \\
      
    % \bottomrule 
    \end{tabular*}
    \caption{Comparing Inference on HC200 dataset. \textit{From left to right}: Gamma corrected source image, 3D LUT Local trained on LocHDR, G-SemTMO trained on LocHDR, G-SemTMO trained on HC200 and Ground truth. HyAB colour distance and PSNR metric scores show significant improvement when G-SemTMO is trained over the style and contrast-specific HC200 images.}
    \label{fig:hc200_result}
\end{figure*}

\subsection{Training \& Inference}

We train over the 200 high contrast images using K-fold cross validation~\cite{friedman2001elements} ($K=4$) with a training-validation data split of $150/50$. We use the ADAMW solver~\cite{adamw} for optimization, weight decay of $5e-4$ and a scheduled learning rate of $10^{-3}$ between epoch $0-150$, $10^{-4}$ after $150^{th}$ epoch and finally $10^{-5}$ after epoch 300.

We trained three networks separately to observe the influence of graph convolutions and style-specific training data:
\begin{itemize}
    \item[-] a network with local semantic information without graph convolutions, Local LUT trained over LocHDR dataset.
    \item[-] a network with local semantic information with graph convolutions, G-SemTMO trained over LocHDR dataset.
    \item[-] a network with local semantic information with graph convolutions, G-SemTMO trained over style-specific HC200 dataset.
\end{itemize}

\begin{figure}[]
\centering
  \includegraphics[width=.485\linewidth]{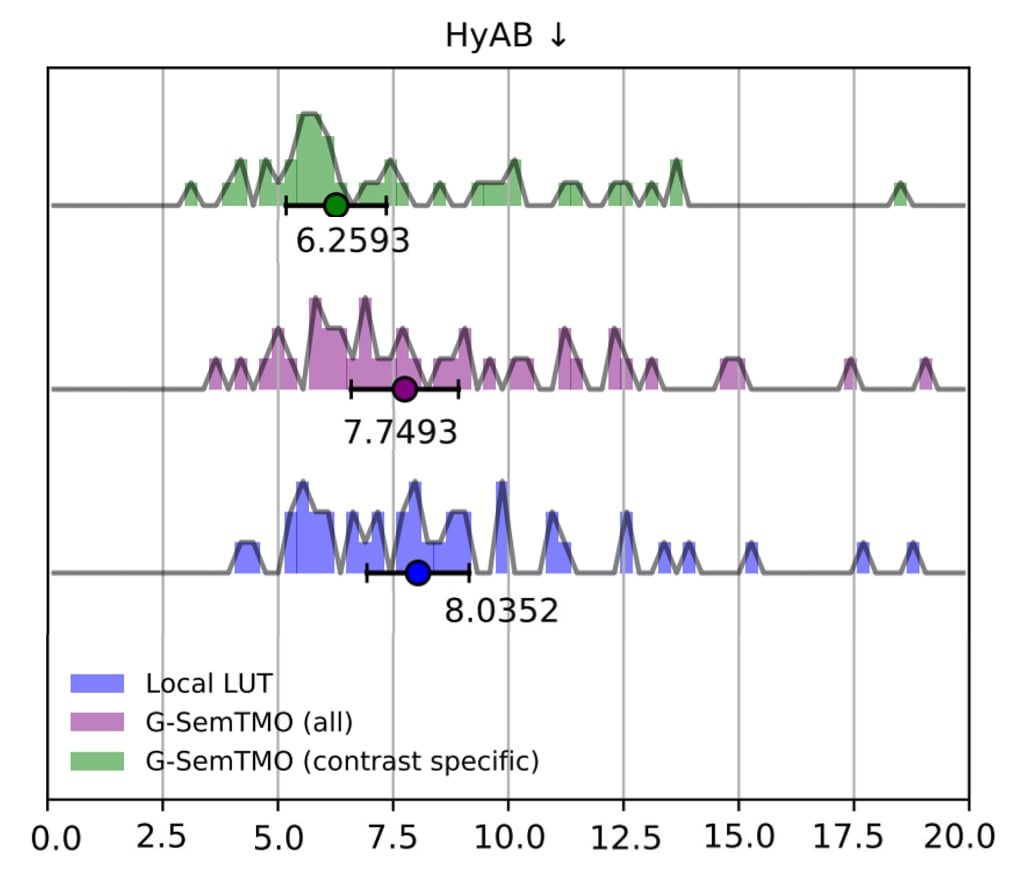}
  \includegraphics[width=.50\linewidth]{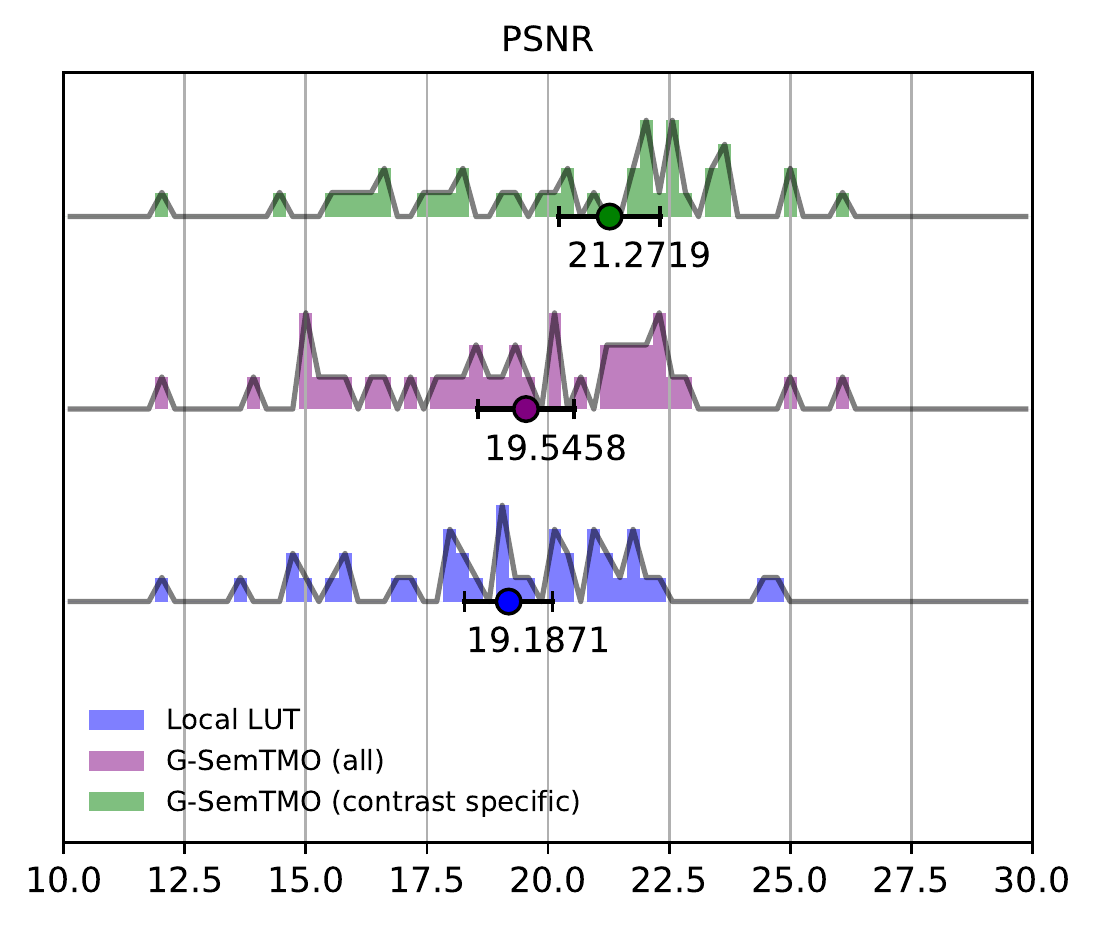}
  \caption{Comparing HyAB and PSNR scores for test images inferred by G-SemTMO trained only on HC200,  G-SemTMO trained on LocHDR and 3D Local LUT trained on LocHDR. HyAB and PSNR histograms show respective scores over 40 test images common to both validation sets. We observe that G-SemTMO trained on the specific style of HC200 produces significantly better inference and colour closeness than the other two networks.}
  \label{fig:hc200_metrics}
\end{figure}

For inference, 40 common images are chosen from the inference sets of LocHDR and HC200. \figref{hc200_result} shows the inference of selected test images for subjective assessment. As mentioned previously, we consider three networks to compare the inference subjectively. For each image in the figure, we see marked improvement in the inference quality of G-SemTMO when trained over style-specific HC200. Network trained over HC200 manages to predict the local contrast in the images closest to the ground truth. This confirms that it is essential for neural networks to be trained on image data with consistent enhancement styles to learn local enhancements better. Previously in \figref{style differentiate}, we have shown that G-SemTMO could learn the different retouching styles made by 5 expert photographers on the FiveK~\cite{fivek} dataset. The training on HC200 shows that for datasets with local adjustments the network can be finetuned by training for specific style variations inside the dataset.
\figref{hc200_metrics} compares the three trained networks objectively on the basis of PSNR and HyAB colour closeness. The HyAB and PSNR histograms for the three networks are plotted along with their median score with a confidence interval of $95\%$. From our experiments, we observe that by training over the entire LocHDR dataset G-SemTMO can only marginally improve over the quality of Local LUT. However, from \figref{hc200_metrics}, we observe significant improvement in G-SemTMO inference quality when trained over a specific style. 

\section{Discussion and limitations}
In our work, we introduced LocHDR, a locally tone mapped dataset of HDR images manually retouched by an expert and G-SemTMO, a novel local tone mapping operator, which can learn global and local tonal transformations from semantic-graph representations of images and the spatial arrangement of the semantic regions.

We compare the results obtained using G-SemTMO in our experiments and ablation studies to the ones tone mapped by the selected reference TMOs and we can confidently claim that graph-based learning can better incorporate semantic awareness in a TMO. We evaluate G-SemTMO on two use-cases. First, we show that G-SemTMO can learn global enhancements from MIT Adobe FiveK dataset~\cite{fivek} better than selected classical and data-driven TMOs. Second, following our novel dataset of locally tone mapped HDR images we show that G-SemTMO can learn local enhancements by letting the graph convolutional network leverage the spatial arrangement of semantic regions. 
When comparing over data from MIT FiveK dataset, our results show that our network can produce images closer to the versions manually retouched by expert photographer E than the other methods. When comparing over data from the LocHDR and HC200 dataset, we observe that the presence of graph convolutions help even further in learning local enhancements with consistent tonal modifications in the training image set in comparison to networks without graph convolutions. 

However, in the process of developing G-SemTMO, we identify some limitations as well. First, our algorithm is reliant on the semantic segmentation of the images to create a graph of their spatial arrangement of the segments. We observe several cases where the label annotations are improper. Image \textit{a1824} in \figref{semantic-awareness} contains a segment \textit{city}, which should clearly belong to the segment \textit{water}.
The improper labels are more challenging with fine grained semantic labelling. Merging labels to coarser segments helps reduce improper annotation to an extent but we still feel the need for a segmentation algorithm and annotated dataset with labels fit for the use case of photography. This can reduce not just improper labelling but also introduce labels that are closer to an expert photographers' impression of a scene.

Second, G-SemTMO in its current state treats all the neighbor semantic segments equally while predicting the latent semantic hints. However, in many cases, semantic segments occupy low percentage of pixels. Image \textit{a5000} in \figref{tc_explain} (bottom) has a very small proportion of pixels annotated as \textit{human}. However, it impacts the tonal adjustment of its neighbor label \textit{vegetation} equally as the label \textit{sky}. One approach to address this would be to have edge-weighted learning where the GCN not only takes the edge adjacency but also the edge importance into account based on how large the semantic segment is.

% \begin{figure}[]
% \centering
%   \includegraphics[width=\linewidth]{Illustrations/tc_compare.jpg}
%   \caption{\textit{\textbf{Comparing inference images with similar `Human Subject' tone curves.}} 3D Local LUT infers tone curves for the human semantic segment (images \textit{a1994} and \textit{a2243}) which are quite similar with a PCM distance of $0.66$. The human semantic region has different semantic neighbours in the two images. Hence, G-SemTMO predicts significantly different tone curves for the same segment with a PCM distance of $6.0513$. }
%   \label{tc_compare}
% \end{figure}

\begin{figure}[]
\centering
  \includegraphics[width=.85\linewidth]{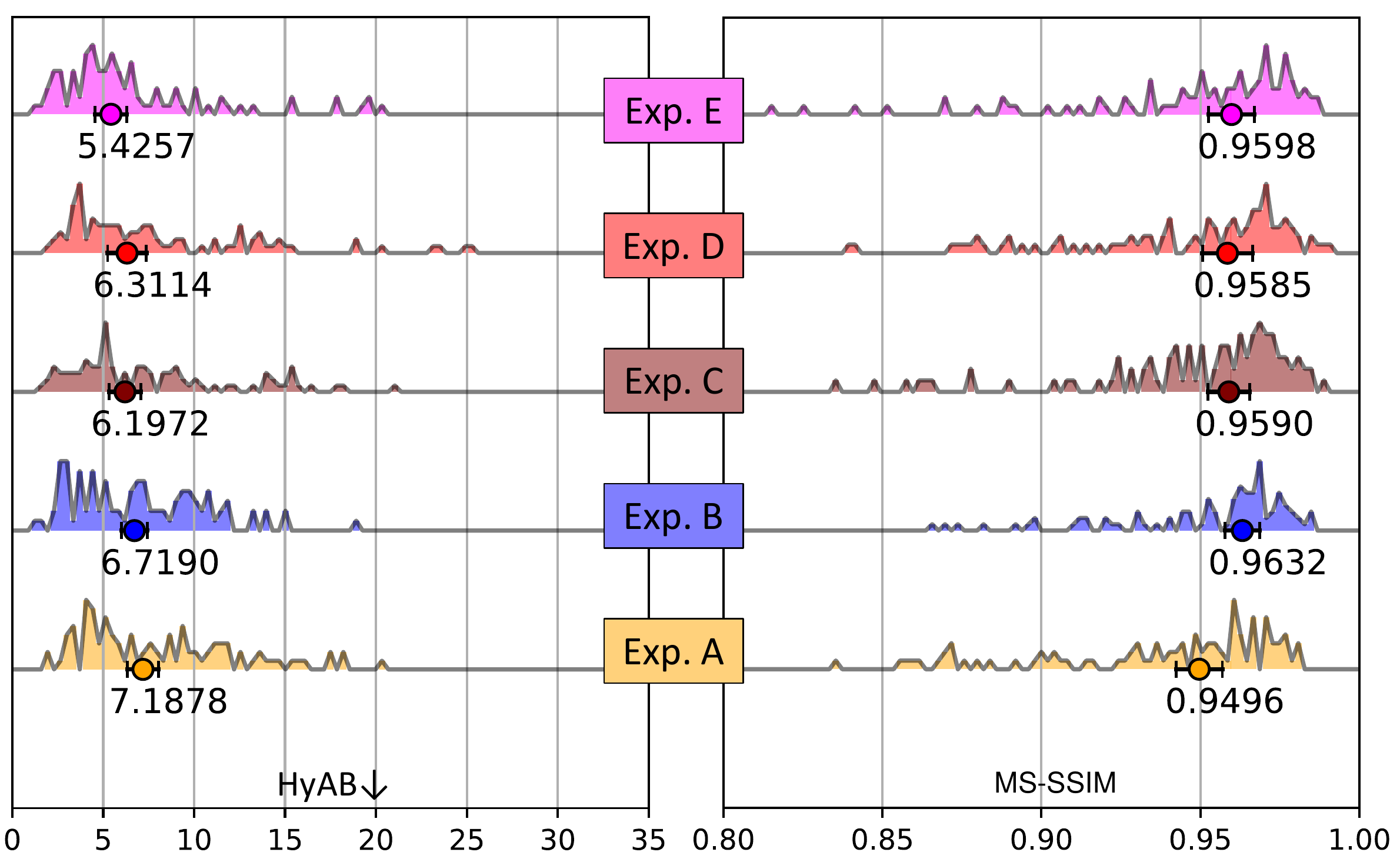}
  \caption{\textit{Learning tonal transformation vs structural attributes}: HyAB metric scores (left) and MS-SSIM scores (right) for network trained over expert E. The network learns tonal transformation specific to the expert. However, there is no discernible difference on the structural similarity scale between networked trained on expert E and others.}
  \label{fig:no_struct}
\end{figure}

Furthermore, while learning global enhancements we observe that each of our networks trained over 5 experts learn tone mapping specific to the expert but do not learn any distinct structural modifications. \figref{no_struct} shows that network trained on expert E learns a colour transformation significantly different from others (validated by HyAB distance) but the MS-SSIM scores do not show a discernible difference. The reason is two-fold: the experts from FiveK dataset do not modify structural parameters such as image sharpness, and although the local contrast parameters may change due to tonal transformation, it does not explicitly modify structural similarity in our result. G-SemTMO does not perform local structural modifications (e.g. sharpening) and it focuses instead on color and contrast transformations. Additionally, it must be noted that our network trains on input images with as-shot camera white balance. So, it is unable to reproduce occasional custom white balance modifications made by the expert. 

Our LocHDR and HC200 locally tone mapped image datasets provide a valuable contribution for development of data-driven TMOs. G-SemTMO has shown that it can learn local enhancements from the dataset and get closer to ground truth compared to networks without graph convolutions. It can learn better when trained over data with specific style. However, we acknowledge that LocHDR has limitations with inconsistency in enhancements and representation of node permutations owing to relatively low number of images. We look forward to improving the dataset with larger number of tone mapped image pairs with more consistent tonal adjustment and wider representation of semantic graphs.

We do not perform a formal subjective comparison of the results as our goal is to produce results that are close to those of an expert photographer rather than to produce the most preferred results. We find the existing full-reference objective metric sufficient for evaluation of that goal. However, we use our network trained on expert~E to tone map images from the Fairchild dataset to show that our results can be generalised for other datasets to produce aesthetically pleasing results too.

\section*{Acknowledgments}
This work has been supported by the European Union's Horizon 2020 research and innovation programme under the Marie Sklodowska-Curie Grant Agreement No. 765911 (RealVision ITN). We would also like to extend our gratitude to Ishani Jayawardhana GJKIU for the hours of dedicated image retouching which helped us create our dataset of images.

\bibliographystyle{IEEEtran}
\bibliography{Biblio.tex}

\newpage

 

% \end{IEEEbiography}



\vfill

\end{document}

%% file: macros.tex
\newcommand{\figref}[1]{Figure~\ref{fig:#1}}
\newcommand{\secref}[1]{Section~\ref{sec:#1}}

\renewcommand{\eqref}[1]{Eq.~\originaleqref{eq:#1}}